\documentclass[onecolumn,12pt,preprint]{aastex62}
\usepackage{amsmath,booktabs,graphicx,subfigure,amssymb,latexsym,graphics,epsfig,exscale,url,booktabs}
\usepackage[flushleft]{threeparttable}
\usepackage{epstopdf}
\usepackage{gensymb}
\usepackage[section]{placeins}
\usepackage[colorinlistoftodos]{todonotes}
\usepackage{float}
\usepackage{chngcntr}
\counterwithin*{footnote}{section}
\usepackage[english]{babel}
\usepackage[utf8x]{inputenc}
\usepackage{rotating}
\begin{document}

\title{The Twisted Magnetic Field of the Protobinary L483}
\correspondingauthor{Erin G. Cox}
\email{erin.cox@northwestern.edu}

\author[0000-0002-5216-8062]{Erin G. Cox}
\affil{Center for Interdisciplinary Exploration and Research in Astronomy (CIERA), Northwestern University, 1800 Sherman Avenue,  
Evanston, IL 60208, USA}

\author[0000-0003-1288-2656]{Giles Novak}
\affil{Center for Interdisciplinary Exploration and Research in Astronomy (CIERA), Northwestern University, 1800 Sherman Avenue,  
Evanston, IL 60208, USA}
\affil{Department of Physics \& Astronomy, Northwestern University, 2145 Sheridan Road, Evanston, IL 60208, USA}

\author{Sarah Sadavoy}
\affil{Department for Physics, Engineering Physics and Astrophysics, Queen’s University, Kingston, ON K7L 3N6, Canada}

\author[0000-0002-4540-6587]{Leslie W. Looney}
\affil{Department of Astronomy, University of Illinois, 1002 West Green Street, Urbana, IL 61801, USA}

\author[0000-0002-3455-1826]{Dennis Lee}
\affil{Center for Interdisciplinary Exploration and Research in Astronomy (CIERA), Northwestern University, 1800 Sherman Avenue,  
Evanston, IL 60208, USA}
\affil{Department of Physics \& Astronomy, Northwestern University, 2145 Sheridan Road, Evanston, IL 60208, USA}

\author{Marc Berthoud}
\affil{Center for Interdisciplinary Exploration and Research in Astronomy (CIERA), Northwestern University, 1800 Sherman Avenue,  
Evanston, IL 60208, USA}
\affil{Engineering + Technical Support Group, University of Chicago, Chicago, IL 60637, USA}

\author{Tyler L. Bourke}
\affil{SKA Observatory, Jodrell Bank, Lower Withington, Macclesfield, Cheshire SK11 9FT, UK}

\author{Simon Coud{\'e}}
\affil{SOFIA Science Center, Universities Space Research Association, NASA Ames Research Center, Moffett Field, California 94035, USA}

\author{Frankie Encalada}
\affil{Department of Astronomy, University of Illinois, 1002 West Green Street, Urbana, IL 61801, USA}

\author{Laura M. Fissel}
\affil{Department for Physics, Engineering Physics and Astrophysics, Queen’s University, Kingston, ON K7L 3N6, Canada}

\author{Rachel Harrison}
\affil{Department of Astronomy, University of Illinois, 1002 West Green Street, Urbana, IL 61801, USA}

\author{Martin Houde}
\affil{Department of Physics and Astronomy, The University of Western Ontario, 1151 Richmond Street, London, Ontario N6A 3K7, Canada}

\author{Zhi-Yun Li}
\affil{Department of Astronomy, University of Virginia, Charlottesville, VA 22904, USA}

\author{Philip C. Myers}
\affil{Center for Astrophysics | Harvard and Smithsonian (CfA), Cambridge, MA 02138, USA}

\author[0000-0002-8557-3582]{Kate Pattle}
\affil{Department of Physics and Astronomy,
University College, Gower St., London WC1E 6BT, UK}

\author{Fabio P. Santos}
\affil{Max Planck Institute for Astronomy, K\"onigstuhl 17, 69117 Heidelberg, Germany
}

\author{Ian W. Stephens}
\affil{Department of Earth, Environment, and Physics, Worcester State University, Worcester, MA 01602, USA}

\author{Hailin Wang}
\affil{Center for Interdisciplinary Exploration and Research in Astronomy (CIERA), Northwestern University, 1800 Sherman Avenue,  
Evanston, IL 60208, USA}
\affil{Department of Physics \& Astronomy, Northwestern University, 2145 Sheridan Road, Evanston, IL 60208, USA}

\author{Sebastian Wolf}
\affil{University of Kiel, Institute of Theoretical Physics and Astrophysics, Leibnizstrasse 15, 24118 Kiel, Germany}

\begin{abstract}
We present H-band (1.65 $\mu$m) and SOFIA HAWC+ 154 $\mu$m polarization observations of the low-mass core L483. Our H-band observations reveal a magnetic field that is overwhelmingly in the E-W direction, which is approximately parallel to the bipolar outflow that is observed in scattered IR light and in single-dish $^{12}$CO observations. From our 154 $\mu$m data, we infer a $\sim$ 45$^{\circ}$ twist in the magnetic field within the inner 5$\arcsec$ (1000 au) of L483. We compare these new observations with published single-dish 350 $\mu$m polarimetry and find that the 10,000 au scale H-band data match the smaller scale 350 $\mu$m data, indicating that the collapse of L483 is magnetically regulated on these larger scales. We also present high-resolution 1.3 mm ALMA data of L483 which reveals it is a close binary star with a separation of 34 au. The plane of the binary of L483 is observed to be approximately parallel to the twisted field in the inner 1000 au. Comparing this result to the $\sim$ 1000 au protostellar envelope, we find that the envelope is roughly perpendicular to the 1000 au HAWC+ field.
Using the data presented, we speculate that L483 initially formed as a wide binary and the companion star migrated to its current position, causing an extreme shift in angular momentum thereby producing the twisted magnetic field morphology observed. More observations are needed to further test this scenario. 
\end{abstract}

\section{Introduction}  \label{sec:intro} 

Star formation begins in molecular clouds and ends in a stellar system, which spans over ten orders of magnitude in spatial scale and in density. These clouds can be quite crowded \citep{mckee2007}, which can obfuscate which dynamics are important. To mitigate this confusion, we can observe Bok globules \citep{bok1947}, which are low-mass, star-forming cores, isolated from the larger molecular cloud and relatively simple in their structure \citep[e.g.,][]{launhardt2013}. The isolated core Lynds 483 \citep{lynds1962}, commonly known as L483, is a well-studied globule located at a distance of $\sim$200 pc \citep{dame1985}. L483 was associated with the Aquila Rift region \citep[distance recently revised to 436 $\pm$ 9 pc][]{ortiz2018}, but \text{Gaia} Data Release 2 astrometry indicates L483 is indeed located at $\sim$ 200 pc \citep{jacobsen2019}, making it a truly isolated region.

The core of L483 hosts the Class 0 protostar IRAS 18148-0440 \citep{parker1988,fuller1995}, though there is some evidence from NIR observations that this object might be transitioning into Class I \citep{tafalla2000}. \citet{jorgensen2004} found evidence of a 1 M$_{\odot}$ central object using velocity gradients of molecular lines, while \citet{oya2017} found the central mass to be 0.1-0.2 M$_{\odot}$ using a ballistic model. \citet{shirley2000} used SCUBA sub-mm observations to show that L483 has an extended structure perpendicular to its outflow axis on core-size ($\sim$ 10,000 au), differing from other low-mass protostars. Molecular line observations of HCN, CS and N$_{2}$H$^{+}$ at 1000 au show a chemically rich infalling envelope surrounding the protostar \citep{jorgensen2004}. At $\sim$ 10,000 au scales, \citet{tobin2010} found a complex 8 $\mu$m extinction morphology surrounding L483 and having an irregular shape. Subsequent sub-mm observations \citep{leung2016} have found a flattened envelope down to $\sim$ 800 au. \citet{leung2016} also used CS(6-7) observations to model the infalling material of L483. \citet{oya2017} used CS observations from ALMA to characterize the infall envelope, and determined that the $\sim$ 800 au structure is rotating. Additionally, the authors found L483 hosts a hot corino \citep{oya2017}. Multiple studies have searched for a disk in this source but have yet to find one down to $\sim$ 20 au resolution \citep[e.g.,][]{oya2017,jacobsen2019}. 

Magnetic fields can 
affect star formation on all spatial scales. In the classical picture of magnetized star formation, the alignment of the angular momentum axis with the magnetic field of the core is expected to be parallel if the field is strong \citep{mouschovias1979}. Conversely, magnetohydrodynamic (MHD) simulations with a misaligned field have shown the collapse process to be more efficient at removing the angular momentum from infalling material thereby allowing disk formation \citep[e.g.,][]{joos2012}. To further study this outflow-field alignment, \citet{chenostriker} examined $\sim$ 100 cores formed in MHD simulations and found that systems where the magnetic field dominated, or those with low turbulence, were more often aligned to the angular momentum axis of the system than those systems where turbulence was dominant.

Dust polarization observations are the most common way to infer the magnetic field morphology in star-forming regions. In the presence of a magnetic field, elongated dust grains are aligned with the field via radiative alignment torques \citep[e.g.,][]{lazarian2007,andersson2015} and can be used to map the morphology of the magnetic field. In star-forming regions, near-infrared polarimetry traces polarization by dichroic extinction of background stars \citep[see, e.g.,][]{alves14}. In this case the linearly polarized light observed is parallel to the magnetic field. This is in contrast to the thermal dust emission at longer wavelengths which directly probes the collapsing region. In this case, the linearly polarized emission from aligned dust grains is perpendicular to the field lines. 

To test whether star formation is magnetically regulated, polarization observations at varying scales have aimed to measure the angle between the magnetic field and outflow \citep[e.g.,][]{davidson2011,stephens13,hull2014,cox2018,galametz2018,galametz2020,sadavoy2018b,sadavoy2018c}. Work done by \citet{hull2014} found that on $\sim$ 1000 au scales the alignment between the field and outflow axis \citep[a proxy for angular momentum direction,][]{pudritz1983} is essentially random. Follow-up work by \citet{hull2019} compiled known interferometric polarization observations of this outflow-field alignment and found the same result, indicating that on 1000 au scales the field is less important. On these envelope size scales, other dynamical processes, such as outflows, can distort the field morphology \citep[e.g.,][]{li2014,hull2017b}. To mitigate these small-scale dynamical distortions, \citet{yen2021} used the JCMT to measure the outflow-field alignment of 62 cores at $\sim$ 0.05-0.5pc. These authors found a mean 3D outflow-field alignment angle of 50$^{\circ}$, thus showing magnetic regulation does not work in all clouds. Conversely, however, the work done by \citet{davidson2011} and \citet{chapman2013} showed some evidence of preferential outflow-field alignment in cores that are isolated \citep[see also ][]{mignonrisse2021}. 

A difference in the outflow-field alignment between isolated cores and non-isolated will have implications for how the natal environment contributes to how easily cores can fragment, if alignment can be linked to the strength of the field \citep{chenostriker}. 
\citet{hennebelle2011} found that a stronger magnetic field can hinder fragmentation on envelope scales. Conversely, others \citep[e.g.,][]{boss2000ApJ} have found fragmentation can increase with a stronger field. Observational work done by \citet{galametz2018,galametz2020} used polarization data from 20 protostars using the SMA and found that magnetic fields in envelopes were either aligned in low angular momentum sources or misaligned in sources with high angular momentum. Additionally, \citet{hull2020} used ALMA polarization to search for hints of how the outflow-field alignment in the binary BHR 71 affected its formation and found an hourglass morphology in one star but evidence of compression of the magnetic field along the outflow of the other star. While these results show hints at how the magnetic field affects protostellar collapse, currently there is not a consensus. Observational studies aimed at understanding the role of magnetic fields and core fragmentation reveal a complex relationship and more work is necessary to link them with theoretical predictions. 

In this paper, we present core-scale magnetic field observations of the isolated core L483 using near-infrared and far-infrared observations. HAWC+ \citep{harper2018} is a polarimeter installed on the Stratospheric Observatory for Infrared Astronomy (SOFIA). It operates in four discrete bands between 50 - 240 $\mu$m. In this paper, we use 154 $\mu$m dust polarization observations from SOFIA/HAWC+ to infer the magnetic field of L483 on envelope scales. We compare the morphology seen in these data with the morphology seen on larger core scales from the Pico dos Dias Observatory using H-band polarimetry. We also compare these data to the CSO-Sharp 350 $\mu$m polarization maps \citep{chapman2013}. Additionally, we use \textit{Herschel}\footnote{\textit {Herschel} is an ESA space observatory with science instruments provided by European-led Principal Investigator consortia and with important participation from NASA.} data to investigate the column density and temperature of the environment surrounding L483. We also use high-resolution ALMA data to determine the binarity of the protostellar system.

This paper is organized as follows. In Section \ref{sec:obs}, we discuss our observations and data reduction methods. In Section \ref{sec:res}, we show our results including polarization maps, and the high-resolution ALMA observations. We discuss the implications of our findings in Section \ref{sec:dis}. We summarize all of this in Section \ref{sec:sum}.

\section{Observations and Data Reduction}  \label{sec:obs}
\subsection{HAWC+ Data}
Observations of L483 were taken using the HAWC+ instrument on SOFIA in Band D (154 $\mu$m) and Band E (214 $\mu$m) 
as part of the project 07\_0184 (PI: Sarah Sadavoy). Band E observations were taken on 2019-07-22 and 2019-09-07 for a total exposure time of 1.116 hours with a FWHM of $\sim$18.2$\arcsec$. The data from the first flight was flagged as off-nominal, and are not included in this analysis. Additionally, the Band E data from the second flight did not show a meaningful polarization signal (see $\chi^{2}$ discussion below), so we show the continuum data only (see Appendix \ref{app:bande}). 
Band D observations were taken on 2019-07-23 and 2019-07-24 for a total of 1.015 hours and with a 
FWHM of $\sim$ 13.6$\arcsec$. Both Band D and Band E data were taken using the standard matched-chop-nod procedure as outlined by \citet{hildebrand2000}. The observations used a chopping frequency of 10.2 Hz, a chop angle of 10$^{\circ}$ (from north increasing east), and a chop amplitude of 250$\arcsec$ in order to avoid including extended emission in the off-positions. These specific chop values were chosen to reduce the low level source flux in our reference beam. 
The observations used a dither pattern, essentially pointing the telescope at four independent locations in a square with an offset of 20$\arcsec$ in Band D and an offset of 27$\arcsec$ in Band E.

We reduced the HAWC+ data manually using the data reduction pipeline as described in \cite{santos2019}. Briefly, the data is first demodulated, and any flagged or bad data is thrown out. This step also accounts for the chopping nature of the observations. The data is then flat fielded to calibrate variations in gain between pixels and to remove data from dead or noisy pixels. The signals reflected and transmitted from the polarizer are then differenced and summed to create Stokes I, Q, and U maps at each half wave plate position. These maps are then flux corrected by combining the fluxes from various nod positions. We then apply atmospheric corrections to these maps using standard models. Finally, the individual observations are combined into a final I, Q, and U map.

We utilize a $\chi^{2}$ analysis on our data to check for internal consistency in our I, Q, and U maps. Briefly, we divide the data into bins and compute I, Q, and U, as well as their corresponding uncertainties, for each bin. We then compare these maps with each other to ensure each is within the claimed errors, i.e., by checking to see if the value of $\chi^{2} \sim 1$ where, 
$\chi^{2}$ = [(actual scatter)/(expected scatter)]$^{2}$. If the value of $\chi^{2} > 1$, then we inflate the error-bars by the square root of the reduced $\chi^{2}$ \citep[for more details see, e.g.,][]{novak2011,chapman2013}. For the L483 dataset, we implemented a more robust version of this inflation by parameterizing our $\chi^{2}$. We sort the pixels according to intensity and then fit the corresponding $\chi^{2}$ values to an expression giving $\chi^{2}$ as a function of intensity. We then inflate each pixel's errors according to that pixel's intensity. 
This in-depth correction was necessary for our dataset due to the low overall total flux of L483.

To address a pointing drift found in the Band D observations, we used a 2D Gaussian fit to the center of emission in each scan and found the instrument pixel corresponding to the peak. This was done prior to the coordinate shift between instrumental and equatorial coordinates. We then manually changed the SIBS value of each scan to reflect the fitted value. The total drift in the pointing was estimated to be $\sim$ 1/3 of the beam ($\sim$ 4\arcsec) and did not significantly change the peak flux measured, or the shape of the source. 

Polarization vectors are calculated using the Polarized Intensity (P), 
shown in equation \ref{eqn:poli} and the Polarized Angle, shown in equation \ref{eqn:pola}. 
\begin{equation}\label{eqn:poli}
    P = \sqrt{Q^{2}+U^{2}-\sigma_{P}^2}
\end{equation}
\begin{equation}\label{eqn:pola}
    \theta = \frac {1}{2} arctan\left( \frac{U}{Q} \right)
\end{equation}
The polarization data is debiased using the most probable estimator in Eq. \ref{eqn:poli} \citep[e.g.,][]{Wardle1974}. Since a negative value of Q or U yields a different measurement of polarization angle, we take the correct quadrant into account when computing $\theta$. The vectors shown are all above a 3$\sigma$ threshold in Polarized Intensity, where the median $\sigma_{P}$ is $\sim$ 0.4 mJy arcsec$^{-2}$. The Nyquist sampled polarization map yields eight detections of polarization for the Band D data and zero detections for the Band E data. The percent polarization is calculated using p = P/I, where I is the total intensity.

\subsection{ALMA Data}
The 1.3 mm (Band 6) ALMA observations of L483 were taken on 2017-08-20 as part of the 2016.1.00085 program (PI: Michael Dunham). The ALMA data have a resolution of $\sim$ 0.09$\arcsec$ and are centered around 225 GHz. The total amount of time on source was $\sim$ 10 minutes. We used the Common Astronomy Software Applications \citep[(CASA),][]{casa} to image the calibrated data from the archive. Figure \ref{fig:alma} was produced using the \texttt{CASA} task \texttt{TCLEAN} with a uniform weighting parameter. Since we used the ALMA data to determine the binarity of L483, we chose a uniform weighting over a natural weighting to prioritize resolution over sensitivity. The final synthesized beam is 0.117$\arcsec$ $\times$  0.079$\arcsec$ at a position angle of \textminus 84$^{\circ}$. We performed one round of self-calibration on this data using a solution interval equal to the scan time. In Appendix \ref{app:alma}, we show these data using briggs weighting.

\subsection{H-band Data} \label{sec:H-band}

The H-band (1.65 $\mu$m) polarimetry data presented here were collected at the Pico dos Dias Observatory\footnote{The Pico dos Dias Observatory is operated by the Brazilian National Laboratory for Astrophysics (LNA), a research institute of the Ministry of Science, Technology and Innovation (MCTI).} in June 2014 using the IAGPOL polarimeter \citep{Magalhaes1996} in combination with a HAWAII 1024$\times$1024 - Cam IV imaging detector.  The instantaneous field of view was 4\arcmin $\times$ 4\arcmin, which with modest dithering enabled us to cover a 4.7\arcmin $\times$ 4.7\arcmin\ field approximately centered on the position of the YSO.   The polarimeter consists of a half-wave plate followed by a Savart analyzer and a spectral filter.  The effect of the analyzer is to produce two orthogonally polarized beams that are imaged side by side on the detector.  Data were collected at 8 half-wave plate positions spaced at 22.5\degree, with each single exposure having a duration of 10 s.  We completed 60 full cycles through each of the 8 half-wave plate positions for a total integration time of approximately 4800 s.

We obtained the stellar polarization fractions $p_H$ and E-vector angles $\theta_H$ via a series of data reduction procedures that included bias and flat-field corrections, sky subtraction, point-source identification, flux measurement, astrometric correction, least square fits to a polarization modulation function, and final calibration using polarimetric standard stars.  Detailed descriptions of this procedure and the calibration data used have been presented in earlier work \citep{Santos2012,Santos2014,Santos2016,Santos2017}. Final uncertainties on the measured polarization fractions were obtained via a quadrature addition of statistical and systematic errors, where the latter were conservatively set equal to 0.1\% based on upper limits for the instrumental polarization calibration error reported by \citet{Santos2012}.  Finally, the measured polarization fractions were debiased in the usual way \citep{Wardle1974,Santos2017} and then measurements not satisfying $p_H>3\sigma_{p_H}$ were rejected. Detections of H-band polarization were obtained for 93 stars.  

\subsection{\textit{Herschel} Data}
To better understand our new data from L483 we also used \textit{Herschel} 250, 350, and 500 $\mu$m \texttt{SPIRE} data\footnote{\texttt{SPIRE} observing labels 1342229186} (Spectral and Photometric Imaging Receiver) and 160 $\mu$m \texttt{PACS} data\footnote{\texttt{PACS} observing labels are 1342228397,1342228398} (Photodetector Array Camera) in our analysis. These data were obtained from the \textit{Herschel} archive\footnote{http://archive.esac.esa.int/hsa/whsa/}. We used these data to give us valuable information on the column density, temperature and optical depth of L483 (see Section \ref{sec:pol}). The \texttt{SPIRE} data were zero-point corrected as described in \citet{sadavoy2018}. These corrected maps were used to create synthetic HAWC+ Band D maps, as described in Appendix \ref{app:fit}. We also use these data to zero-point correct the \texttt{PACS} data using the procedure described in Appendix \ref{app:fit}.

\section{Results}  \label{sec:res}

\subsection{Polarization Results}\label{sec:pol}
\begin{figure*}
\includegraphics[scale = 0.36]{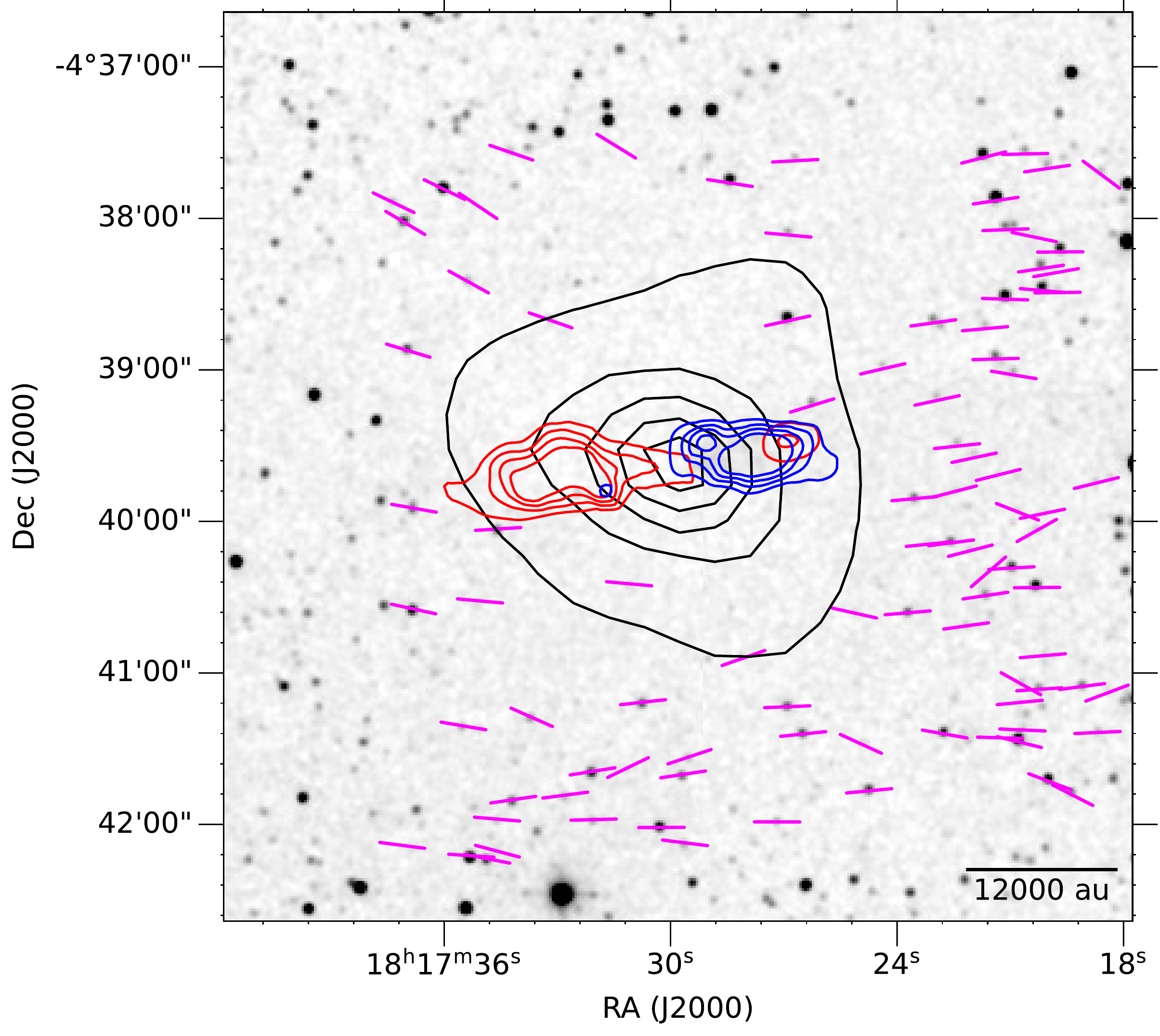}
\includegraphics[scale = 0.67]{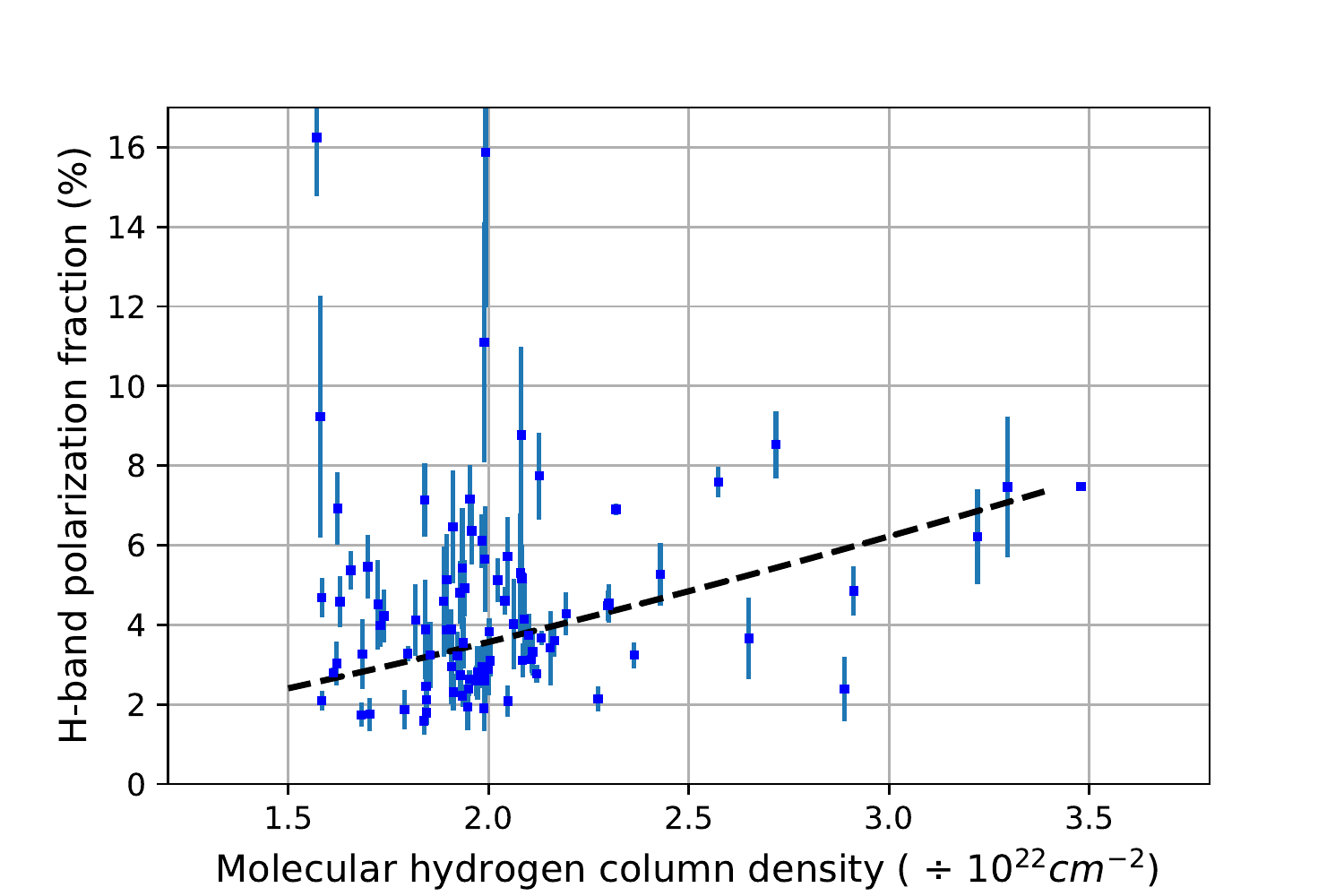}
\caption{\textit{left:} H-band inferred magnetic field shown in magenta vectors. Background image is 2MASS H-band. Black contours are \textit{Herschel} 500 $\mu$m intensity shown at 100, 200, 300, 400, and 500 MJy/sr. $^{12}$CO outflow from IRAM shown in red and cyan contours \citep{tafalla2000} separated by 2.1 K km s$^{-1}$. \textit{right:} Polarization fraction $p_H$ vs.\ column density $N_{H_2}$ for H-band stellar polarization measurements.  Column density is obtained from graybody fits to Herschel maps (see Appendix \ref{app:fit}).  The black dashed line shows a power law fit to the data.  The fitted equation is $ p_H = p_0 \times (N_{H_2})^\alpha $ where $N_{H_2}$ is the column density divided by $10^{22} cm^{-2}$.  The best fit parameters are $p_0 = (1.38 \pm 0.15) \% $ and $\alpha = 1.37 \pm 0.11$.}
\label{fig:hband}
\vspace{0.5cm}
\end{figure*}

In Figure \ref{fig:hband} ({\em left panel}) we show the corresponding vectors for all 93 H-band polarization detections, drawn parallel to $\theta_H$ in order to illustrate the inferred magnetic field angle. (Recall that for polarization by absorption, the inferred magnetic field direction is parallel to the measured E-vector \cite[e.g.,][]{draine1997,lazarian2007}.)  The background image in this figure is 2MASS H-band, and we also superpose the contours of \textit{Herschel} 500 $\mu$m dust emission (black), as well a single-dish IRAM map of the red- and blue-shifted CO outflow lobes \citep{tafalla2000}.  

Given the relatively small distance to the target ($\sim$200 pc, see Section \ref{sec:intro}) and the relatively small sky area mapped ($< 0.1$ pc$^2$ at 200 pc), we expect very few if any foreground stars. We investigated the distances to our H-band stars using the \textit{Gaia} edr3 database \citep{gaia1,gaia2,gaia3} and found that most stars in our sample are too obscured by the globule to be accurately measured by \textit{Gaia}. Out of our 93 H-band stars, 13 have \textit{Gaia} parallax measurements, and only two out of that sample have robust (i.e., $\ge$ 3 $\sigma$) parallax detections. We calculate the distances to these two stars using the procedure outlined in \citet{bailerjones21} and find distances of 880 pc and 1060 pc. Both of these stars are at the edge of the cloud, not towards the center of L483. Due to the dusty nature of the globule obscuring the central stars to \textit{Gaia} observations, we must use a different metric to determine if our H-band data are tracing the magnetic field of L483.

To test the extent to which the measured H-band polarization is due to dust in the globule, rather than being caused by distant material far behind the globule, we used estimates of molecular hydrogen column density $N_{H_2}$ obtained from our fits to the \textit{Herschel} data (see Appendix \ref{app:fit} and Figure \ref{fig:nh2}) to carry out a study of the dependence of $p_H$ on $N_{H_2}$. The right panel of Figure \ref{fig:hband} plots these two quantities for all 92 polarization detections for which estimates of $N_{H_2}$ are available.  (One of the 93 stars having H-band polarization detections was located at positions just off the western edge of our $N_{H_2}$ map.)  Despite the presence at low $N_{H_2}$ of a small minority of outliers having $p_H$ near or above 10\%, we can see an overall tendency for $p_H$ to increase with $N_{H_2}$.  To quantify this trend, we carried out a weighted least squares power law fit to the data.

The fitted equation was $ p_H = p_0 \times (N_{H_2}/10^{22}cm^{-2})^\alpha $.  The weight of each of the 92 data points used in the fit was set equal to the inverse square of the estimated error in $p_H$ for that point.  The best fit power law is shown in Figure \ref{fig:hband} ({\em right panel}) using a black dashed line. From this fit we find $p_0 = (1.38 \pm 0.15) \% $ and the power law exponent is $\alpha = 1.37 \pm 0.11$.  This fit result suggests that the H-band polarization measurements toward higher column densities are indeed tracing the magnetic field in the globule itself \citep[e.g.,][]{goodman1995,chapman2011}.  We conclude that the magnetic field in the outer regions of the globule, where the dust extinction is small enough to see through at H-band, appears to have an East-West orientation - approximately parallel to the orientation of the CO outflow \citep[e.g.,][]{fuller1995,tafalla2000,velusamy14}.

To determine whether our HAWC+ 154 $\mu$m polarization results for L483 are due to emission or absorption, we calculate the optical depth ($\tau$) at different points in the cloud using our fitted values for column density and temperature obtained from the Herschel data (see Appendix \ref{app:fit}). At the central pixel, i.e., the flux peak, and at 23.9$\arcsec$ resolution, we find a value of $\tau_{154\mu m} = 0.037$ using the 160 $\mu$m data. 
This value of $\tau_{154\mu m}$ shows that the far-infrared emission of L483 is optically thin, even at the peak. Therefore, our HAWC+ polarization observations are tracing the magnetic field morphology from emission, instead of absorption, as we are seeing in the near-infrared, H-band observations. \citet{zielinski2021} also found their HAWC+ polarization data to be due to emission in the Bok globule B335.

\begin{figure*}
\includegraphics[scale = 0.33]{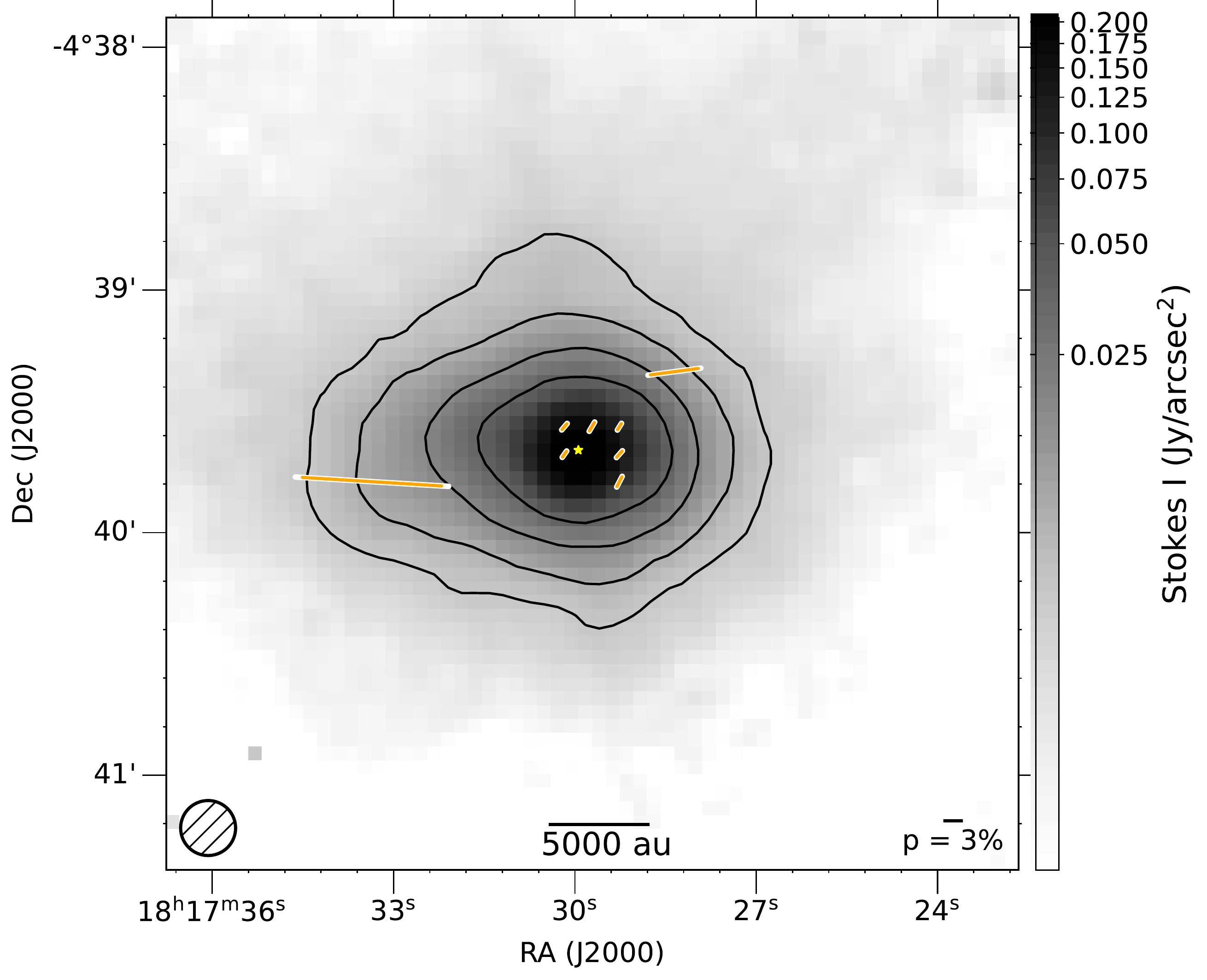}
\includegraphics[scale = 0.33]{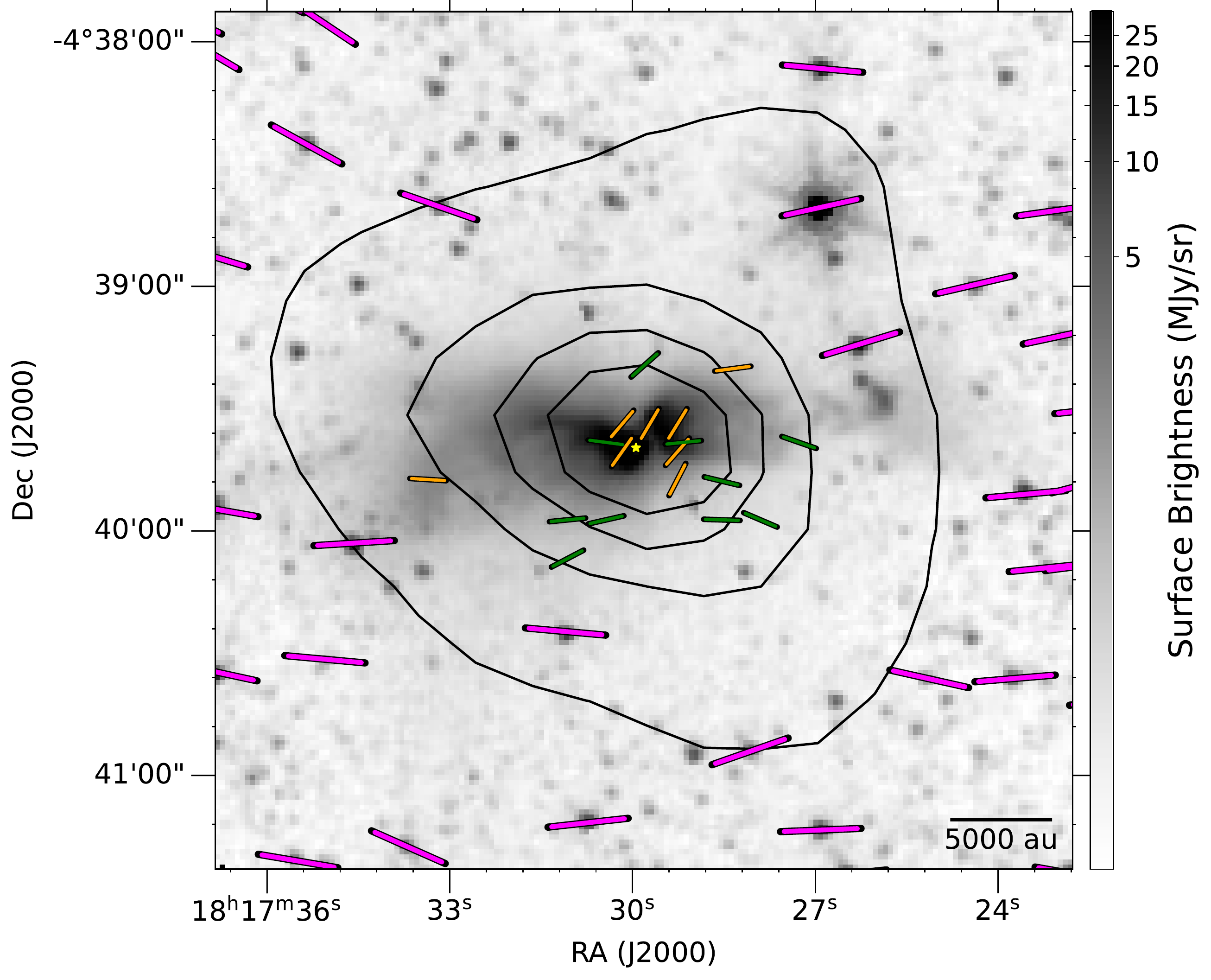}
\caption{\textit{left:} SOFIA Band D (154 $\mu$m) observations of L483. The 154 $\mu$m total intensity emission is shown in grayscale and contours. The black contours are shown at 3, 5, 10, and 20$\sigma$, $\sigma$ = 1.9 mJy arcsec$^{-2}$. Scaled magnetic field vectors in orange with the polarization percentage scalebar shown on the bottom right. The inner vectors have a polarization $\gtrsim$ 1\% while the two outer vectors are 7\% and 20.4\%.  The resolution of the observations is $\sim$ 13.6$\arcsec$ and is shown on the bottom left.
\textit{right:} A subset of the H-band vectors (magenta) shown on the \textit{Spitzer} 4.5 $\mu$m background image. SHARP 350 $\mu$m \citep{chapman2013} vectors are shown in green and HAWC+ 154 $\mu$m vectors are shown in orange. Note that the HAWC+ vectors are Nyquist sampled while the SHARP vectors correspond independent beams. All vectors shown are inferred magnetic field orientation.
Both panels show the location of the ALMA protostar as a yellow star.}
\label{fig:sofial483}
\end{figure*}

Figure \ref{fig:sofial483} shows the inferred magnetic field revealed by SOFIA Band D observations, as well as the recorded dust polarization percentages. The vectors shown have been rotated by 90$^{\circ}$ since they are tracing polarized emission \citep[e.g.,][]{lazarian2007,andersson2015}.
This is shown superimposed on the 154 $\mu$m total intensity. The peak intensity at 154 $\mu$m is 262 mJy/arcsec$^{2}$, and the contours show the source shape starting at 3$\sigma$,
where $\sigma$ is the sensitivity of the image which is $\sim$ 1.9 mJy/arcsec$^{2}$. We note that the use of debiased vectors compared to non-debiased does not change the morphology of the field nor the number of vectors detected, and has only a minor effect on the polarization fraction of the vectors (the polarization fractions for debiased vectors are 95 - 97\% of those for corresponding non-debiased vectors).
These observations show an E-W field orientation in the outer regions of the core, and a $\sim$ 45$^{\circ}$ counter-clockwise twist (relative to the E-W field) towards the central source. This figure shows that the inferred field in this compact region remains organized. 

The right panel of Figure \ref{fig:sofial483} shows the inferred magnetic field using H-band data (magenta vectors), CSO/SHARP 350 $\mu$m data \citep[green,][]{chapman2013}, and the SOFIA 154 $\mu$m data (orange vectors). The vectors are superimposed on a \textit{Spitzer} 4.5 $\mu$m map of L483 \citep{velusamy14}. The H-band vectors show a clear E-W magnetic field direction, which is approximately parallel to the outflow seen in the \textit{Spitzer} map (see also CO outflow lobes in Figure \ref{fig:hband}). Since the H-band polarimetry corresponds to stars behind the cloud, each H-band vector is seen on top of an individual star. The 350 $\mu$m vectors are also seen to have a mostly E-W direction. 
The 154 $\mu$m magnetic field vectors are shown in orange in this figure to compare with the larger field, which are at a significantly different position angle. We discuss these three datasets, including the overall E-W field direction and the counter-clockwise twist, in Section \ref{sec:morph}. An important feature to note is that while the 154 $\mu$m data show Nyquist sampled vectors (4 vectors per resolution element), the 350 $\mu$m data is showing one vector per resolution element, which is how \citet{chapman2013} reported the data.

\subsection{Total Intensity Results}\label{sec:cont}
\begin{figure*}
\includegraphics[scale = 0.37]{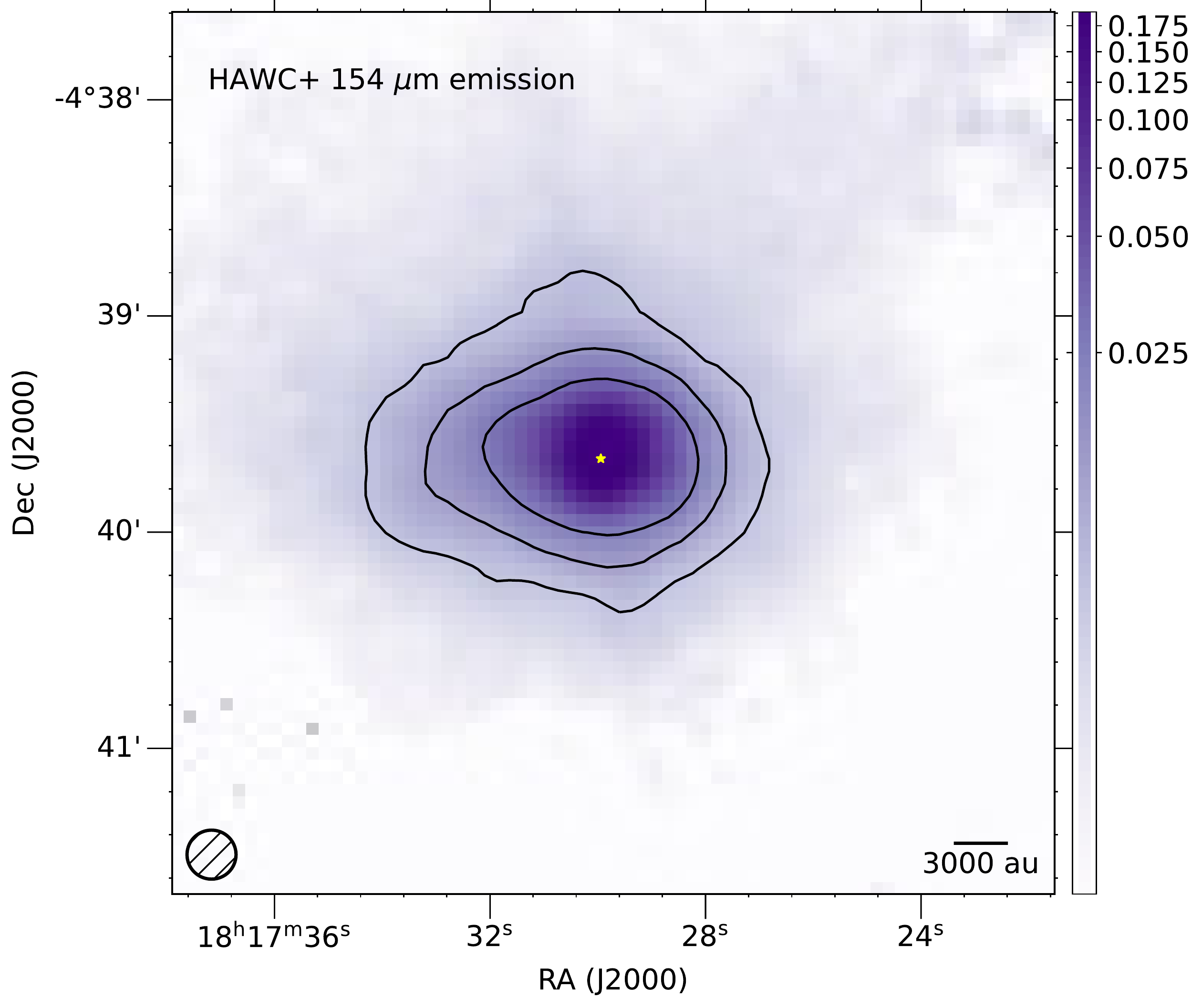}
\includegraphics[scale = 0.37]{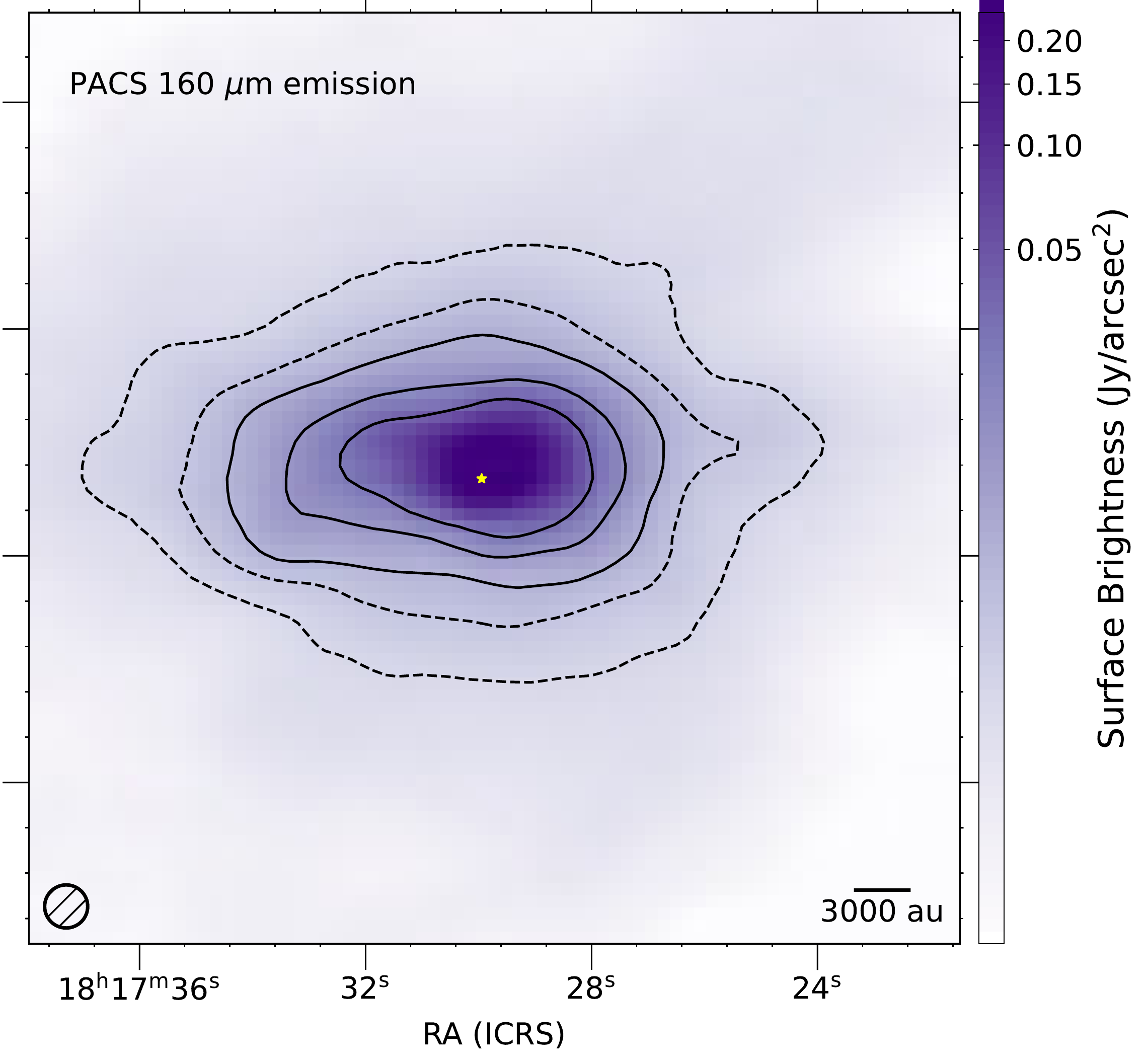}
\caption{\textit{left:} HAWC+ 154 $\mu$m total intensity shown in the colorscale and contours. The solid black contours shown correspond to 6, 12, and 25 mJy arcsec$^{-2}$. The $\sim$13.6$\arcsec$ beam is shown in the bottom left. \textit{right: PACS} 160 $\mu$m L483 data shown in the colorscale and contours. The \texttt{PACS} $\sim$11.4$\arcsec$ beam is shown in the bottom left. We show dashed black contours corresponding to 2.5 and 4 mJy arcsec$^{-2}$, and solid black contours at 6, 12, 25 mJy arcsec$^{-2}$. To compare the source shape between the two telescopes we plot the same contours on the HAWC+ data (6, 12, 25 mJy arcsec$^{-2}$), excluding the two levels which are lower than the sensitivity of the HAWC+ data. The shape of L483 is elongated in the N-S direction in the 154 $\mu$m data compared to the 160 $\mu$m data. This is expected due to the nature of chop-nod observations. In both images we show the location of the ALMA protostar with a yellow star.}
\label{fig:pacs}
\end{figure*}

We show our zero point corrected (zpc) \texttt{PACS} 160 $\mu$m map of L483 in Figure \ref{fig:pacs}. The peak intensity measured in this map is 429 mJy arcsec$^{-2}$, and the source is elongated in the E-W direction. This elongation is in the same direction as the observed outflow \citep[e.g.,][]{tafalla2000,velusamy14}. The 160 $\mu$m data do not show as much extended emission of L483 as the \texttt{SPIRE} maps do \citep{sadavoy2018}. The \texttt{PACS} maps exhibit a surface brightness that is low compared to that of the galaxy, suggesting the temperature of L483 is cold and therefore should have less emission at 160 $\mu$m. 

\begin{figure}
\includegraphics[scale = 0.33]{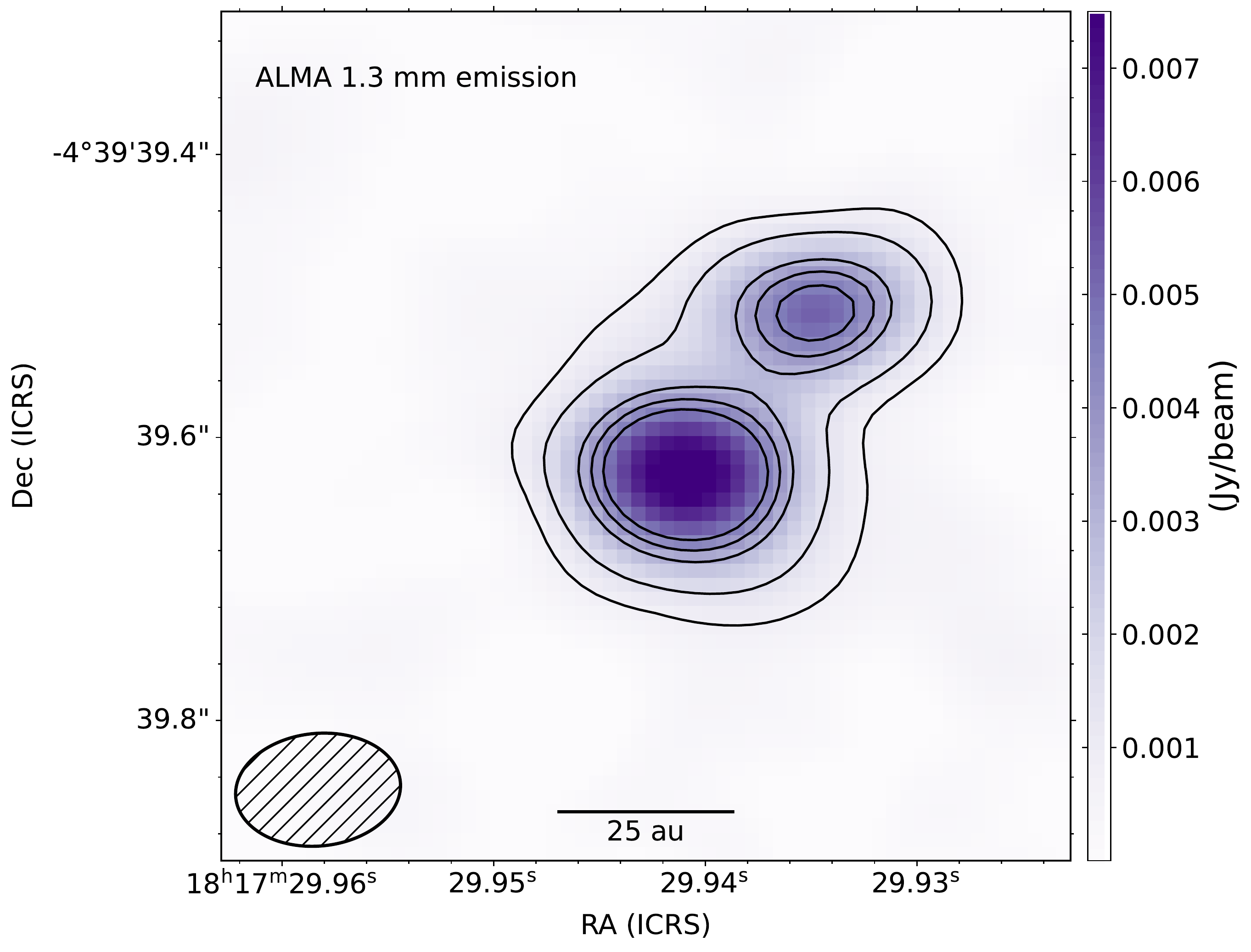}
\caption{Band 6 ALMA data of L483 revealing that L483 hosts a binary star system. This image prioritized resolution over sensitivity using uniform weighting in the cleaning process. We show black contours at 5, 10, 20, 25, 30$\sigma$, $\sigma$ = 0.15 mJy/beam. The ALMA beam is shown in the bottom left, and a scale bar of 25 au is shown in the bottom center.}
\label{fig:alma}
\end{figure}

In Figure \ref{fig:alma} we show the 0.09$\arcsec$ resolution ALMA 1.3 mm observations of L483. These observations reveal, for the first time, that L483 is a binary system. The brighter, southern source has a peak flux of $\sim$ 8 mJy/beam, while the dimmer source in the north-west direction has a peak flux of $\sim$ 5 mJy/beam. Each star looks like a point-source, suggesting they harbor a small disk. This may indicate that magnetic braking is an important effect in this system. At a distance of 200 pc, the binary has a separation of $\sim$ 34 au. \citet{galametz2020} argue that L483 is a binary with a wider ($\sim$ 400 au) separation based on a 5$\sigma$ detection at $\sim$ 0.65 mJy/beam \citep{oya2017} at a resolution of 0.4$\arcsec$. The high-resolution observations reported in this paper filter out this structure, and we posit that this was an observation of fluffy envelope material surrounding L483.

Table \ref{tab:flux} summarizes results of total intensity maps from \texttt{PACS} (160 $\mu$m), HAWC+ (154 $\mu$m), and ALMA (1.3 mm). For each map, we list coordinates of intensity peaks, peak intensity values, peak fluxes, as well as map angular resolution and map sensitivity. The HAWC+ 154 $\mu$m reported peak flux is $\sim$87\% of the  \textit{Herschel} \texttt{PACS} 160 $\mu$m reported peak flux. From the HAWC+ observers guidebook and \citet{gordon2018}, we expect our calibration to be within 10\% of the true value. There are a few reasons that could be contributing to this continuum flux discrepancy. First, this could be explained if there is a slight error in the total calibration. Additionally, the difference between the two peak flux values may be explained by the difference in beam size used for a source that is quite peaked in its emission. Finally, it is possible this discrepancy can be accounted for due to the slight difference in wavelength between the two datasets. We also note that since the HAWC+ observations are taken using chop-nod, L483 appears to be smeared out in the direction of the chop\footnote{This beam smearing is likely due to the lack of bright guide stars near the target combined with a 250$\arcsec$ chop amplitude, the maximum value allowed for HAWC+ observing.} 
(the 154 $\mu$m continuum emission in left panel of  Figure \ref{fig:pacs} is more rounded than the 160 $\mu$m emission in right panel of Figure \ref{fig:pacs}). 
This smearing of L483 in the N-S direction is likely due to the large chop amplitude of the secondary mirror, which was necessary to get a clean atmospheric background subtraction. It is also possible that the smearing is due to a slightly larger beam size due to the pointing drift described in Section \ref{sec:obs}. 
Nevertheless, we do not expect the polarization results to be affected by any calibration discrepancies due to our $\chi^{2}$ analysis. 

\begin{sidewaystable}
\begin{center}
\caption{Positions and Fluxes of Intensity Peaks in L483}
\vspace{0.2cm}
\begin{tabular}{l l l c c c c}
\hline
\hline

 &  RA (J2000) & Dec (J2000) & Intensity & Peak Flux & Beam ($\arcsec$)& Sensitivity\\ \hline
HAWC+ 154 $\mu$m & 18:17:29.89 & -4:39:39.78 &  262 mJy/arcsec$^{2}$ &54.68 Jy/beam& 13.6 & 1.9 mJy/arcsec$^{2}$\\
\texttt{PACS} 160 $\mu$m & 18:17:29.93 & -4:39:39.45 & 429 mJy/arcsec$^{2}$ & 62.97 Jy/beam &11.4 & 0.61 mJy/arcsec$^{2}$\\

\hline \hline
ALMA 1.3 mm \\
\hline
IRAS 18148-0440A & 18:17:29.9428 & -4:39:39.599 & & 8.3 mJy/beam & 0.09 & 0.15 mJy/beam\\
IRAS 18148-0440B & 18:17:29.9365 & -4:39:39.483 & & 5.3 mJy/beam & 0.09 & 0.15 mJy/beam
\\
\hline
\end{tabular}  
\end{center}
\scriptsize Position, intensity, peak flux, beam size and sensitivity of L483 for various observations. Beam area is calculated using 1.13*(FWHM)$^{2}$.
\label{tab:flux}
\end{sidewaystable}

\section{Discussion}  \label{sec:dis}
\subsection{Magnetic Field Morphology of L483}\label{sec:morph}
Measuring the magnetic field morphology across spatial scales can inform our study of magnetically regulated collapse in star formation. 
The \textit{Planck}\footnote{Based on observations obtained with Planck (http://www.esa.int/Planck), an ESA science mission with instruments and contributions directly funded by ESA Member States, NASA, and Canada.} all-sky survey mapped the L483 region. Using the \textit{Planck} Legacy Archive, we obtained the 850 $\mu$m Stokes I, Q, and U parameters at the location of L483 at 5$\arcmin$ resolution. We find that the debiased polarization percentage is 0.71\% with an uncertainty of 0.55\%, indicating that there was not a robust detection of polarization at 5$\arcmin$. The inferred magnetic field angle measurement from these observations is 70$^{\circ}$ with an uncertainty of $\sim$ 17$^{\circ}$, which would indicate that on these large scales the magnetic field of L483 has an E-W orientation. Due to the lack of a robust detection from these observations, we merely acknowledge that the field morphology on $\sim$ 0.3 pc scales may be in roughly the same direction as our H-band data (see Section \ref{sec:H-band} and Figure \ref{fig:hband}). 

Our H-band data reveal a mostly E-W magnetic field direction in L483. The field probed by these observations is larger than the core-structure of L483 traced by the \textit{Herschel} 500 $\mu$m emission. Towards the central near-IR emission, some of these data start to look as though they may be pinched towards the protostar, yet they still mostly persist in the E-W direction. In the right panel of Figure \ref{fig:sofial483}, we show the CSO/SHARP 350 $\mu$m inferred magnetic field in green \citep{chapman2013}. These SHARP observations probe down to $\sim$ 2000 au and are, on average, in the same E-W direction as the H-band vectors. This consistency of parallelism with the outflow across these spatial scales 
suggests the core of L483 is likely an example of magnetically regulated collapse, as seen in B335 \citep{maury2018} and L1448N \citep{kwon2019}. This result is amplified by the high degree of order in the H-band vectors, which can indicate a strong magnetic field \citep{mocz2017}. Thus, if the magnetic field on core scales is strong, and we see this degree of alignment with the outflow, we do not expect large-scale turbulence to dominate the dynamical processes of L483. This is in line with the isolation of this cloud.

Our 154 $\mu$m data reveal a twist in the inferred magnetic field direction in the central emission of L483 (note the six vectors closest to the protostar in Figure \ref{fig:sofial483}). The outer two vectors are in the E-W direction and agree well with the H-band data and the SHARP data. This twist is $\sim$ 45$^{\circ}$ counter-clockwise with-respect-to the larger-scale field. The twist in the central inferred magnetic field may be due to small-scale dynamics changing the direction of the field, such as rotation or binary interaction. Previous studies have compared field morphology across scales, with some evidence for abrupt changes in field orientation \citep[e.g.,][]{hull2017b,pattle2021}. Some models also show similar evidence of such an abrupt change \citep[e.g.,][]{myers2020}.

Although the resolution of the HAWC+ data is $\sim$ 2700 au, it is possible the central, twisted vectors are probing a hot, small central source. Dust models used in \citet{jacobsen2019} (see their Figure 9) show that the dust emission at 154 $\mu$m is likely centrally concentrated - confined to a 5$\arcsec$ region, corresponding to 1000 au. Therefore we suggest that our central HAWC+ vectors are from dust emission also on this small size scale. If the polarized emission is, in fact, from the central 1000 au, then this may resolve the apparent discrepancy between the SHARP 350 $\mu$m vectors, which has a resolution of $\sim$ 2000 au, and the 154 $\mu$m vectors (see right panel Figure \ref{fig:sofial483}).

\subsection{How the Field is Affecting the Formation of L483}

Using MHD simulations, \citet{chenostriker} analyzed 100 protostellar cores and found that the degree of misalignment between the angular momentum axis of the core with respect to the magnetic field direction increases when turbulence increases, or when the magnetic field is weak. Thus, a system with its field aligned with its angular momentum could be dominated by magnetic field energy. To determine the angular momentum direction of L483, we use its outflow \citep{pudritz1983}. As shown in Figure \ref{fig:hband}, the large-scale CO outflow, which was observed using the 30 m single dish IRAM \citep{tafalla2000}, aligns approximately parallel to the E-W direction and extends approximately $\sim$ 10,000 au in either direction of the central source. The field lines extend further in the E-W direction in the core of L483 than the outflow, as well as N and S of the outflow. This outflow extends to approximately the outer contour of the \textit{Herschel} 500 $\mu$m emission. 

In the classical picture of magnetized star formation, a dominant magnetic field is aligned with the angular momentum of the system \citep{mouschovias1979}. In this scenario, the infalling material efficiently loses angular momentum via magnetic braking. This effect, however, was shown to be too efficient in that it suppressed disk growth in most systems, and nonideal MHD effects were proposed as a means to reduce the efficiency  \citep[e.g.,][]{li2011}. More recent results have shown magnetic braking is less efficient when the axes are misaligned \citep[e.g.,][]{joos2012}, thus allowing disks to grow \citep[e.g.,][]{li2014b}. The ALMA data presented in Section \ref{sec:cont} do not seem to exhibit large disks-- the stars are unresolved so the disks appear to be less than 18 au. Such a small disk size may indicate that magnetic braking has been important in the formation of L483.

Surveys such as TADPOL \citep{hull2014} have searched for observational confirmation of this outflow-field (mis)alignment, and found that at $\sim$ 1000 au scales the alignment is random. A recent result from the BISTRO survey \citep{yen2021} showed that in 62 cores the mean 3D magnetic field is $\sim$ 50$^{\circ}$ misaligned with respect to the outflows. We note that this survey did find the projected outflow-field alignment to be within 15-30$^{\circ}$. Nevertheless, these large surveys show that there is no strong preference in outflow-field alignment. The results of \citet{chapman2013} disagree with these other surveys as they found evidence of aligned fields in seven isolated cores. Therefore it is possible the degree of isolation of the target could dictate how well aligned it is.

The core-scale field is approximately parallel to the outflow seen in L483. Though it is possible for the outflow to have affected the field direction on small scales, it is unlikely for it to have been able to affect this field direction we see on such a large scale since the outflow energy becomes low compared to the gravitational energy density of the cloud. Therefore, we posit our data are showing that the magnetic field influenced the formation of this outflow. We do note that, when considering the results of, e.g., \citet{hull2014,hull2019,yen2021}, it is possible that the alignment of the magnetic field of L483 and its outflow axis is by chance. 

As material collapses from the cloud down to the central protostar, it tends to be directed by the field direction \citep{allen2003}. This is seen in clouds using an Histogram of Relative Orientation analysis \citep[see, e.g., ][]{soler2013,lee2021}, and is thought to continue down to smaller scales. While the change in morphology from large to small scales can indicate a change in the small-scale dynamics, it is possible that the field is still controlling the flow of the material. If that is the case, then our observations are in line with MHD simulations \citep[e.g.,][]{allen2003} predicting magnetically regulated collapse of the protostar. Meaning the infalling material of L483 is flowing along the field lines and accumulating onto the observed flattened envelope. This has been seen in polarization observations using JCMT in the Ophiuchus cloud \citep{pattle2021}. On the other hand, with the inclusion of more non-ideal MHD effects in simulations, it has been shown that the magnetic field might play an important, yet more subtle role in the collapse due to these effects impeding the field's ability to control gas kinematics \citep[e.g.,][]{zhao2018}. 

It is possible that at some point during the protostellar collapse the magnetic field will be shaped by dynamical processes such as outflows and winds \citep{davidson2011,hull2017b}. Furthermore, for sufficiently high densities charged dust grains may disappear causing the field to decouple from the collapsing material \citep{zhao2018}. With the 154 $\mu$m results shown in this paper, we see evidence of continuity in the outer two vectors, but it is unclear if the twisted field is continuous with the larger scale field, or if, alternatively, the structure has somehow been ripped apart. Since the process of accretion can be sporadic, it is plausible that nonisotropic accretion could have some sort of torque on the inner structure, thus rotating the observed field. \citet{oya2018} report an asymmetry in the observed molecular emission in the envelope of L483, which may be further evidence of such an event.

\subsection{L483 Flattened Infall Envelope and Binarity}
\begin{figure}
\includegraphics[scale = 0.37]{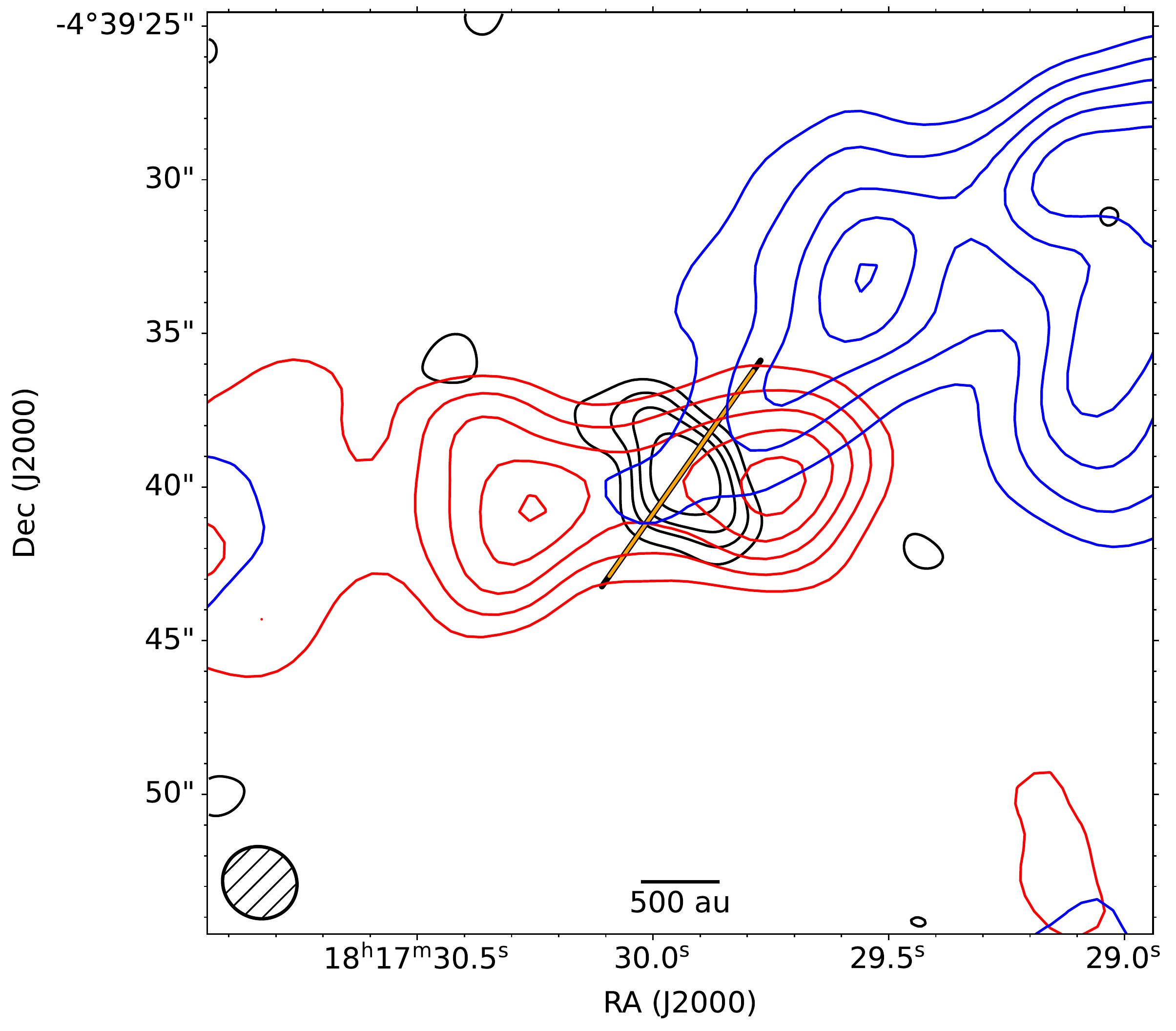}
\caption{Flattened infall envelope of L483 shown in black contours from 870 $\mu$m SMA observations \citep{leung2016}. The beam of the SMA data ($\sim$ 2.35$\arcsec$) is shown in the bottom left corner. Contours are shown at 3, 5, 7, 10$\sigma$, where $\sigma$ is 3.4 mJy beam$^{-1}$. The outflow at 4$\arcsec$ resolution in $^{12}$CO(1-0) from OVRO is seen in the red and blue contours \citep{velusamy14}. The red and blue contours are spaced 0.4 Jy beam$^{-1}$. The unweighted mean of six measurements of field direction at 154 $\mu$m is shown in the orange vector.}
\label{fig:avg}
\end{figure}

\citet{leung2016} found evidence for a flattened envelope of $\sim$ 1000 au size using SMA 850 $\mu$m data as seen in Figure \ref{fig:avg}. This SMA map suggests that this structure is rotated with-respect-to the the large-scale outflow by $\sim$ 30$^{\circ}$. This is reminiscent of our observation that the field here is also rotated with respect to the outflow, which may be a coincidence. Taking the outflow orientation to be 100$^{\circ}$ (see below) and relying on the estimate of 30$^{\circ}$ from \citet{leung2016}, we obtain a smaller counter-clockwise rotation of just 20$^{\circ}$. Nevertheless, it is interesting that this structure is approximately perpendicular to the magnetic field traced by HAWC+. If the collapse of L483 is magnetically regulated on these scales, then we expect to observe a flattened structure perpendicular to the field lines \citep[e.g.,][]{allen2003}. Though the field probed at these scales has shifted from the large scale, it remains organized in its morphology.

The envelope magnetic field morphology of L483 is reminiscent of the \textit{Planck} results for the highest densities probed in ten molecular clouds, in that the field is seen to be perpendicular to the elongated structure \citep{planck2016}. Others have explored the relationship between elongated/filamentary structures and magnetic fields at densities higher than \textit{Planck} can probe and have found a variety of results. \citet{monsch2018}, \citet{sadavoy2018c}, and \citet{pillai2020} have all found evidence for parallel fields in dense filaments. However, in all three cases the parallel fields were not found at the main central concentration of mass. In fact, \citet{pillai2020} found the field to be mostly perpendicular to the elongation at the main concentration of mass, similar to our L483 results. This is consistent with perpendicular fields observed by \citet{chapman2013} in other flattened envelopes seen with perpendicular fields \citep[see also,][]{allen2003}.

The collapse of the protostar and the fragmentation of the core are intimately related to the outflow of the system, and current outflow observations in L483 appear to show a consistent direction. Using HCO$^{+}$ observations, \citet{park2000} found the outflow in L483 has a position angle (PA) of 95$^{\circ}$. The large-scale $^{12}$CO (2-1) observations \citep{tafalla2000} and the outer lobes of the smaller-scale $^{12}$CO (1-0) observations \citep{velusamy14} nicely trace the scattered light 4.5 $\mu$m outflow (see, Figure \ref{fig:sofial483}) with a PA of 105$^{\circ}$. In Figure \ref{fig:avg}, we show the average direction of the 154 $\mu$m magnetic field with the $^{12}$CO (1-0) observations from \citet{velusamy14} and the SMA 850 $\mu$m infall envelope \citep{leung2016}. While we cannot rule out the possibility that the magnetic field we are observing is being compressed along the outflow cavity \citep[as was seen by e.g.,][]{davidson2011,hull2017b}, we do not find convincing evidence that this is what the data are showing. 

At 1000 au scales, the infalling envelope complicates the kinematics and the line emission no longer shows a strong preference for $\sim$ 100$^{\circ}$ (see Figure \ref{fig:avg}). 
In fact, observations of CS(7-6) in L483 are modeled as infalling material along the outflow walls \citep{leung2016}. Combined with high-resolution molecular observations \citep[e.g.,][]{oya2017,jacobsen2019}, it is plausible that these dynamical processes have a role in the observed twist in the magnetic field at 1000 au. While the large-scale magnetic field of L483 appears to be consistent with being strongly magnetized (i.e., the magnetic energy is larger than the turbulent energy), the complexity of the small-scale outflow indicates the magnetic field may not be dominant in the formation of its protostar. A change in angular momentum or gas infall direction on small scales could be causing the twist in the field seen in the infall envelope.

The ALMA data shown here reveal that there are at least two stars forming in L483 (Figure \ref{fig:alma}). The projected binary separation is $\sim$ 30 au. Interestingly, this separation of the binary is parallel to the inner magnetic field traced at 154 $\mu$m. It is unclear from our observation alone whether this is a random alignment or if the symmetry axis of where the binary forms with-respect-to the field direction is important in its formation. We note that while the shape of the ALMA beam is elliptical, it is not elongated in the direction of the binary star and therefore we do not expect this discovery to be a result of beam smearing.

The close binary of L483 brings up questions regarding its formation. Its natal magnetic field on 10,000 au scales is well-ordered and likely quite strong which should dampen fragmentation on envelope scales \citep{zhao2018}. However, it is possible that the companion star initially formed at a larger distance and migrated inward via the accretion of material that has undergone strong magnetic braking \citep{zhaoli2013}. This is in line with simulations by \citet{offner2016} who find an offset between the orbital plane of some close binaries and their outflows. While this is not uncommon in observations \citep{tobin2016}, \citet{offner2016} argue that these systems may have formed via turbulent fragmentation followed by migration to small separations, as opposed to gravitational instabilities in a disk. This formation scenario might be why the magnetic field of L483 at 1000 au is misaligned with its outflow, which was also seen by \citet{galametz2020} using SMA 870 $\mu$m polarimetry data. Our observed field morphology differs by $\sim$ 45$^{\circ}$ from that of the SMA, however those observations are lacking in sensitivity and only produced one magnetic field vector. What is clear from these observations is that there is a dynamical process affecting the magnetic field morphology at the $\sim$ 1000 au scale.

\subsection{Comparison to L1157}
In this section we compare the morphology we observe in L483 to that observed in another well-studied, low-mass protostar, L1157. L1157 is located in the Cepheus flare at a distance of $\sim$ 340 pc \citep{sharma2020} and, like L483, shows evidence of a flattened structure from $\sim$ 10,000 au \citep{looney2007} down to $\sim$ 1000 au scales \citep{kwon2019}.
\citet{shirley2000} observed both L483 and L1157 at 850 $\mu$m with SCUBA and found at 15$\arcsec$ ($\sim$ few thousand au) scales the two differed in their extended structures such that L483's intensity distribution is oriented perpendicular to its outflow, while L1157's is extended parallel to its outflow. The large-scale magnetic field of L1157 is seen to be parallel to its outflow \citep{chapman2013}, similar to what we see in L483. On smaller scales, the field direction in L1157 remains continuous in this direction \citep{stephens13,hull2014}, in contrast to the twist seen in L483. L1157 exhibits an hourglass field morphology on envelope scales \citep{stephens13,hull2014} while L483 does not. Both L1157 and L483 do not show signs of a Keplerian disk down to 10-20 au scales \citep{tobin2013,jacobsen2019}. Like L483, L1157 is a close ($\sim$ 16 au) binary \citep{tobin2022}; however the plane of its binary is perpendicular to the direction of the field. The plane of the binary in L483 is parallel to the direction of the 1000 au scale magnetic field. We suggest that in both sources the magnetic field controlled the large scale collapse and in L1157 the field is still strong enough to regulate the collapse on small scales. However, in L483 the geometry of the field compared to the plane of its binary suggests that this change in angular momentum due to the possible migration process was strong enough to influence the magnetic field morphology. More studies of inner envelope-scale magnetic fields in close binaries are necessary to determine if this configuration of field direction and binary plane is common, and, if so, the reason for the alignment.

\section{Summary}  \label{sec:sum}

In this paper, we present new far-IR (154 $\mu$m) polarization results from SOFIA/HAWC+ in the Bok globule L483. We also show new H-band polarization observations of this region. Additionally, we present new ALMA 1.3 mm total intensity observations of L483. Our main results are:

\begin{enumerate}
    \item We find organized dust polarization around L483. We also find the direction of polarization changes close to the central protostar. The HAWC+ 154 $\mu$m inferred field shows a twist approximately 45$^{\circ}$ counter-clockwise to the larger scale field. This morphology differs from the typical hourglass shape thought to be seen in gravitational collapse \citep[e.g.,][]{girart06, rao09, stephens13}, yet remains organized. This may be indicative of an event that altered the field, such as the formation of another star.
    \item We show H-band polarization data that reveals a magnetic field in the core of L483 having an E-W orientation. This ordered field can be traced to larger distances than the outflow seen in \citet{tafalla2000}, indicating that the magnetic field likely was important in the initial collapse of L483. \citet{chenostriker} show that an outflow parallel to the magnetic field can be indicative of a strongly magnetized cloud. From our H-band data, we argue that the magnetic field is likely dynamically important in the collapse of the core. 
    \item  The change in direction may also indicate that the field is not as strong on small scales. The field appears to be almost orthogonal to the 1000 au-envelope structure found by \citet{leung2016}, and may be funneling material onto this structure as seen in \citet{pattle2021}. Such an occurrence might imply that while the field is not strong enough to resist the change of direction between the scales probed by near-IR and HAWC+ it remains dynamically significant on the 1000 au scale. 
    \item Using 1.3 mm ALMA observations, we show for the first time that L483 is forming at least two stars. The observed binary has a projected separation of $\sim$ 30 au. We show that the plane of the binary appears to be parallel to the magnetic field at 1000 au scales, differing from the situation found in another close binary, L1157. This hints that the formation of the binary, and thus a change in angular momentum, in L483 is responsible for the twist observed in the magnetic field, though more observations are needed to know if this is a common occurrence.
\end{enumerate}

The observations shown in this paper highlight the benefits of using multi-wavelength and multi-scale data to gain insight into the collapse of a protostar. L483 is often regarded as a simple source; yet with high-resolution observations, we reveal it has at least two stars forming. Further theoretical investigation into the magnetic field on these scales is needed to understand how it interacts with gravitational collapse, rotation, and binary formation. 

\clearpage

\acknowledgments
{\centering ACKNOWLEDGMENTS \par}\

Data cubes containing the CO outflow observations for L483 \citep{tafalla2000,velusamy14} and the SMA 850 $\mu$m continuum observations \citep{leung2016} were kindly provided by Mario Tafalla, Thangasamy Velusamy, and Gigi Leung. This work was based [in part] on observations made with the NASA/DLR Stratospheric Observatory for Infrared Astronomy (SOFIA). SOFIA is jointly operated by the Universities Space Research Association, Inc. (USRA), under NASA contract NNA17BF53C, and the Deutsches SOFIA Institut (DSI) under DLR contract 50 OK 2002 to the University of Stuttgart. This paper makes use of the following ALMA data: ADS/JAO.ALMA\#2016.1.00085.S. ALMA is a partnership of ESO (representing its member states), NSF (USA) and NINS (Japan), together with NRC (Canada) and NSC and ASIAA (Taiwan), in cooperation with the Republic of Chile. The Joint ALMA Observatory is operated by ESO, AUI/NRAO and NAOJ. The National Radio Astronomy Observatory is a facility of the National Science Foundation operated under cooperative agreement by Associated Universities, Inc.
PACS has been developed by a consortium of institutes led by MPE (Germany) and including UVIE (Austria); KU Leuven, CSL, IMEC (Belgium); CEA, LAM (France); MPIA (Germany); INAF-IFSI/OAA/OAP/OAT, LENS, SISSA (Italy); IAC (Spain). This development has been supported by the funding agencies BMVIT (Austria), ESA-PRODEX (Belgium), CEA/CNES (France), DLR (Germany), ASI/INAF (Italy), and CICYT/MCYT (Spain). 
SPIRE has been developed by a consortium of institutes led by Cardiff University (UK) and including Univ. Lethbridge (Canada); NAOC (China); CEA, LAM (France); IFSI, Univ. Padua (Italy); IAC (Spain); Stockholm Observatory (Sweden); Imperial College London, RAL, UCL-MSSL, UKATC, Univ. Sussex (UK); and Caltech, JPL, NHSC, Univ. Colorado (USA). This development has been supported by national funding agencies: CSA (Canada); NAOC (China); CEA, CNES, CNRS (France); ASI (Italy); MCINN (Spain); SNSB (Sweden); STFC, UKSA (UK); and NASA (USA). This publication makes use of SPIRE data products \citep{herschel250doi,herschel350doi,herschel500doi}.
This research has made use of the NASA/IPAC Infrared Science Archive, which is funded by the National Aeronautics and Space Administration and operated by the California Institute of Technology. 
This publication also makes use of data products from the Two Micron All Sky Survey \citep{2masspaper}, which is a joint project of the University of Massachusetts and the Infrared Processing and Analysis Center \citep{2massdoi}, funded by the National Aeronautics and Space Administration and the National Science Foundation. 
This publication has made use of data products based on observations obtained with Planck (http://www.esa.int/Planck), an ESA science mission with instruments and contributions directly funded by ESA Member States, NASA, and Canada \citep{planck850doi}. 
This work has made use of data from the European Space Agency (ESA) mission
{\it Gaia} (\url{https://www.cosmos.esa.int/gaia}), processed by the {\it Gaia}
Data Processing and Analysis Consortium (DPAC,
\url{https://www.cosmos.esa.int/web/gaia/dpac/consortium}). Funding for the DPAC
has been provided by national institutions, in particular the institutions
participating in the {\it Gaia} Multilateral Agreement.

We thank the anonymous referee for their helpful comments in improving this manuscript. 
LWL acknowledges support from NSF AST-1910364. ZYL is supported in part by NSF AST-1815784 and NASA 80NSSC20K0533. K.P. is a Royal Society University Research Fellow, supported by grant number URFR1211322. IWS acknowledges support for this work by NASA through award \#08\_0186 issued by USRA.

\facility{IRSA, SOFIA, \it{Herschel}, ALMA, \it{Planck}, \it{Gaia}}
\software{CASA \citep{casa}, Matplotlib \citep{matplotlib},  Numpy \citep{numpy}, Astropy
\citep[http://www.astropy.org; ][]{astropy2013}, 
APLpy \citep{aplpy}, SciPy \citep{scipy}}

\begin{small}
\bibliographystyle{apj}
\bibliography{ms}

\begin{thebibliography}{}
\expandafter\ifx\csname natexlab\endcsname\relax\def\natexlab#1{#1}\fi
\providecommand{\url}[1]{\href{#1}{#1}}
\providecommand{\dodoi}[1]{doi:~\href{http://doi.org/#1}{\nolinkurl{#1}}}
\providecommand{\doeprint}[1]{\href{http://ascl.net/#1}{\nolinkurl{http://ascl.net/#1}}}
\providecommand{\doarXiv}[1]{\href{https://arxiv.org/abs/#1}{\nolinkurl{https://arxiv.org/abs/#1}}}

\bibitem[{{2MASS Team}(2020)}]{2massdoi}
{2MASS Team}. 2020, 2MASS All-Sky Survey Scan Information Table,  IPAC,
  \dodoi{10.26131/IRSA111}

\bibitem[{{Allen} {et~al.}(2003){Allen}, {Li}, \& {Shu}}]{allen2003}
{Allen}, A., {Li}, Z.-Y., \& {Shu}, F.~H. 2003, \apj, 599, 363,
  \dodoi{10.1086/379243}

\bibitem[{{Alves} {et~al.}(2014){Alves}, {Frau}, {Girart}, {Franco}, {Santos},
  \& {Wiesemeyer}}]{alves14}
{Alves}, F.~O., {Frau}, P., {Girart}, J.~M., {et~al.} 2014, \aap, 569, L1,
  \dodoi{10.1051/0004-6361/201424678}

\bibitem[{{Andersson} {et~al.}(2015){Andersson}, {Lazarian}, \&
  {Vaillancourt}}]{andersson2015}
{Andersson}, B.~G., {Lazarian}, A., \& {Vaillancourt}, J.~E. 2015, \araa, 53,
  501, \dodoi{10.1146/annurev-astro-082214-122414}

\bibitem[{{Astropy Collaboration} {et~al.}(2013){Astropy Collaboration},
  {Robitaille}, {Tollerud}, {Greenfield}, {Droettboom}, {Bray}, {Aldcroft},
  {Davis}, {Ginsburg}, {Price-Whelan}, {Kerzendorf}, {Conley}, {Crighton},
  {Barbary}, {Muna}, {Ferguson}, {Grollier}, {Parikh}, {Nair}, {Unther},
  {Deil}, {Woillez}, {Conseil}, {Kramer}, {Turner}, {Singer}, {Fox}, {Weaver},
  {Zabalza}, {Edwards}, {Azalee Bostroem}, {Burke}, {Casey}, {Crawford},
  {Dencheva}, {Ely}, {Jenness}, {Labrie}, {Lim}, {Pierfederici}, {Pontzen},
  {Ptak}, {Refsdal}, {Servillat}, \& {Streicher}}]{astropy2013}
{Astropy Collaboration}, {Robitaille}, T.~P., {Tollerud}, E.~J., {et~al.} 2013,
  \aap, 558, A33, \dodoi{10.1051/0004-6361/201322068}

\bibitem[{{Bailer-Jones} {et~al.}(2021){Bailer-Jones}, {Rybizki}, {Fouesneau},
  {Demleitner}, \& {Andrae}}]{bailerjones21}
{Bailer-Jones}, C.~A.~L., {Rybizki}, J., {Fouesneau}, M., {Demleitner}, M., \&
  {Andrae}, R. 2021, \aj, 161, 147, \dodoi{10.3847/1538-3881/abd806}

\bibitem[{{Bok} \& {Reilly}(1947)}]{bok1947}
{Bok}, B.~J., \& {Reilly}, E.~F. 1947, \apj, 105, 255, \dodoi{10.1086/144901}

\bibitem[{{Boss}(2000)}]{boss2000ApJ}
{Boss}, A.~P. 2000, \apjl, 545, L61, \dodoi{10.1086/317332}

\bibitem[{{Chapman} {et~al.}(2011){Chapman}, {Goldsmith}, {Pineda}, {Clemens},
  {Li}, \& {Kr{\v{c}}o}}]{chapman2011}
{Chapman}, N.~L., {Goldsmith}, P.~F., {Pineda}, J.~L., {et~al.} 2011, \apj,
  741, 21, \dodoi{10.1088/0004-637X/741/1/21}

\bibitem[{{Chapman} {et~al.}(2013){Chapman}, {Davidson}, {Goldsmith}, {Houde},
  {Kwon}, {Li}, {Looney}, {Matthews}, {Matthews}, {Novak}, {Peng},
  {Vaillancourt}, \& {Volgenau}}]{chapman2013}
{Chapman}, N.~L., {Davidson}, J.~A., {Goldsmith}, P.~F., {et~al.} 2013, \apj,
  770, 151, \dodoi{10.1088/0004-637X/770/2/151}

\bibitem[{{Chen} \& {Ostriker}(2018)}]{chenostriker}
{Chen}, C.-Y., \& {Ostriker}, E.~C. 2018, \apj, 865, 34,
  \dodoi{10.3847/1538-4357/aad905}

\bibitem[{{Cox} {et~al.}(2018){Cox}, {Harris}, {Looney}, {Li}, {Yang}, {Tobin},
  \& {Stephens}}]{cox2018}
{Cox}, E.~G., {Harris}, R.~J., {Looney}, L.~W., {et~al.} 2018, \apj, 855, 92,
  \dodoi{10.3847/1538-4357/aaacd2}

\bibitem[{{Dame} \& {Thaddeus}(1985)}]{dame1985}
{Dame}, T.~M., \& {Thaddeus}, P. 1985, \apj, 297, 751, \dodoi{10.1086/163573}

\bibitem[{{Davidson} {et~al.}(2011){Davidson}, {Novak}, {Matthews}, {Matthews},
  {Goldsmith}, {Chapman}, {Volgenau}, {Vaillancourt}, \&
  {Attard}}]{davidson2011}
{Davidson}, J.~A., {Novak}, G., {Matthews}, T.~G., {et~al.} 2011, \apj, 732,
  97, \dodoi{10.1088/0004-637X/732/2/97}

\bibitem[{{Draine} \& {Weingartner}(1997)}]{draine1997}
{Draine}, B.~T., \& {Weingartner}, J.~C. 1997, \apj, 480, 633,
  \dodoi{10.1086/304008}

\bibitem[{{Fuller} {et~al.}(1995){Fuller}, {Lada}, {Masson}, \&
  {Myers}}]{fuller1995}
{Fuller}, G.~A., {Lada}, E.~A., {Masson}, C.~R., \& {Myers}, P.~C. 1995, \apj,
  453, 754, \dodoi{10.1086/176437}

\bibitem[{{Gaia Collaboration} {et~al.}(2016){Gaia Collaboration}, {Prusti},
  {de Bruijne}, {Brown}, {Vallenari}, {Babusiaux}, {Bailer-Jones}, {Bastian},
  {Biermann}, {Evans}, {Eyer}, {Jansen}, {Jordi}, {Klioner}, {Lammers},
  {Lindegren}, {Luri}, {Mignard}, {Milligan}, {Panem}, {Poinsignon},
  {Pourbaix}, {Randich}, {Sarri}, {Sartoretti}, {Siddiqui}, {Soubiran},
  {Valette}, {van Leeuwen}, {Walton}, {Aerts}, {Arenou}, {Cropper}, {Drimmel},
  {H{\o}g}, {Katz}, {Lattanzi}, {O'Mullane}, {Grebel}, {Holland}, {Huc},
  {Passot}, {Bramante}, {Cacciari}, {Casta{\~n}eda}, {Chaoul}, {Cheek}, {De
  Angeli}, {Fabricius}, {Guerra}, {Hern{\'a}ndez}, {Jean-Antoine-Piccolo},
  {Masana}, {Messineo}, {Mowlavi}, {Nienartowicz}, {Ord{\'o}{\~n}ez-Blanco},
  {Panuzzo}, {Portell}, {Richards}, {Riello}, {Seabroke}, {Tanga},
  {Th{\'e}venin}, {Torra}, {Els}, {Gracia-Abril}, {Comoretto},
  {Garcia-Reinaldos}, {Lock}, {Mercier}, {Altmann}, {Andrae}, {Astraatmadja},
  {Bellas-Velidis}, {Benson}, {Berthier}, {Blomme}, {Busso}, {Carry},
  {Cellino}, {Clementini}, {Cowell}, {Creevey}, {Cuypers}, {Davidson}, {De
  Ridder}, {de Torres}, {Delchambre}, {Dell'Oro}, {Ducourant}, {Fr{\'e}mat},
  {Garc{\'\i}a-Torres}, {Gosset}, {Halbwachs}, {Hambly}, {Harrison}, {Hauser},
  {Hestroffer}, {Hodgkin}, {Huckle}, {Hutton}, {Jasniewicz}, {Jordan},
  {Kontizas}, {Korn}, {Lanzafame}, {Manteiga}, {Moitinho}, {Muinonen},
  {Osinde}, {Pancino}, {Pauwels}, {Petit}, {Recio-Blanco}, {Robin}, {Sarro},
  {Siopis}, {Smith}, {Smith}, {Sozzetti}, {Thuillot}, {van Reeven}, {Viala},
  {Abbas}, {Abreu Aramburu}, {Accart}, {Aguado}, {Allan}, {Allasia},
  {Altavilla}, {{\'A}lvarez}, {Alves}, {Anderson}, {Andrei}, {Anglada Varela},
  {Antiche}, {Antoja}, {Ant{\'o}n}, {Arcay}, {Atzei}, {Ayache}, {Bach},
  {Baker}, {Balaguer-N{\'u}{\~n}ez}, {Barache}, {Barata}, {Barbier}, {Barblan},
  {Baroni}, {Barrado y Navascu{\'e}s}, {Barros}, {Barstow}, {Becciani},
  {Bellazzini}, {Bellei}, {Bello Garc{\'\i}a}, {Belokurov}, {Bendjoya},
  {Berihuete}, {Bianchi}, {Bienaym{\'e}}, {Billebaud}, {Blagorodnova},
  {Blanco-Cuaresma}, {Boch}, {Bombrun}, {Borrachero}, {Bouquillon}, {Bourda},
  {Bouy}, {Bragaglia}, {Breddels}, {Brouillet}, {Br{\"u}semeister},
  {Bucciarelli}, {Budnik}, {Burgess}, {Burgon}, {Burlacu}, {Busonero}, {Buzzi},
  {Caffau}, {Cambras}, {Campbell}, {Cancelliere}, {Cantat-Gaudin}, {Carlucci},
  {Carrasco}, {Castellani}, {Charlot}, {Charnas}, {Charvet}, {Chassat},
  {Chiavassa}, {Clotet}, {Cocozza}, {Collins}, {Collins}, {Costigan}, {Crifo},
  {Cross}, {Crosta}, {Crowley}, {Dafonte}, {Damerdji}, {Dapergolas}, {David},
  {David}, {De Cat}, {de Felice}, {de Laverny}, {De Luise}, {De March}, {de
  Martino}, {de Souza}, {Debosscher}, {del Pozo}, {Delbo}, {Delgado},
  {Delgado}, {di Marco}, {Di Matteo}, {Diakite}, {Distefano}, {Dolding}, {Dos
  Anjos}, {Drazinos}, {Dur{\'a}n}, {Dzigan}, {Ecale}, {Edvardsson}, {Enke},
  {Erdmann}, {Escolar}, {Espina}, {Evans}, {Eynard Bontemps}, {Fabre},
  {Fabrizio}, {Faigler}, {Falc{\~a}o}, {Farr{\`a}s Casas}, {Faye}, {Federici},
  {Fedorets}, {Fern{\'a}ndez-Hern{\'a}ndez}, {Fernique}, {Fienga}, {Figueras},
  {Filippi}, {Findeisen}, {Fonti}, {Fouesneau}, {Fraile}, {Fraser}, {Fuchs},
  {Furnell}, {Gai}, {Galleti}, {Galluccio}, {Garabato}, {Garc{\'\i}a-Sedano},
  {Gar{\'e}}, {Garofalo}, {Garralda}, {Gavras}, {Gerssen}, {Geyer}, {Gilmore},
  {Girona}, {Giuffrida}, {Gomes}, {Gonz{\'a}lez-Marcos},
  {Gonz{\'a}lez-N{\'u}{\~n}ez}, {Gonz{\'a}lez-Vidal}, {Granvik}, {Guerrier},
  {Guillout}, {Guiraud}, {G{\'u}rpide}, {Guti{\'e}rrez-S{\'a}nchez}, {Guy},
  {Haigron}, {Hatzidimitriou}, {Haywood}, {Heiter}, {Helmi}, {Hobbs},
  {Hofmann}, {Holl}, {Holland}, {Hunt}, {Hypki}, {Icardi}, {Irwin}, {Jevardat
  de Fombelle}, {Jofr{\'e}}, {Jonker}, {Jorissen}, {Julbe}, {Karampelas},
  {Kochoska}, {Kohley}, {Kolenberg}, {Kontizas}, {Koposov}, {Kordopatis},
  {Koubsky}, {Kowalczyk}, {Krone-Martins}, {Kudryashova}, {Kull}, {Bachchan},
  {Lacoste-Seris}, {Lanza}, {Lavigne}, {Le Poncin-Lafitte}, {Lebreton},
  {Lebzelter}, {Leccia}, {Leclerc}, {Lecoeur-Taibi}, {Lemaitre}, {Lenhardt},
  {Leroux}, {Liao}, {Licata}, {Lindstr{\o}m}, {Lister}, {Livanou}, {Lobel},
  {L{\"o}ffler}, {L{\'o}pez}, {Lopez-Lozano}, {Lorenz}, {Loureiro},
  {MacDonald}, {Magalh{\~a}es Fernandes}, {Managau}, {Mann}, {Mantelet},
  {Marchal}, {Marchant}, {Marconi}, {Marie}, {Marinoni}, {Marrese},
  {Marschalk{\'o}}, {Marshall}, {Mart{\'\i}n-Fleitas}, {Martino}, {Mary},
  {Matijevi{\v{c}}}, {Mazeh}, {McMillan}, {Messina}, {Mestre}, {Michalik},
  {Millar}, {Miranda}, {Molina}, {Molinaro}, {Molinaro}, {Moln{\'a}r},
  {Moniez}, {Montegriffo}, {Monteiro}, {Mor}, {Mora}, {Morbidelli}, {Morel},
  {Morgenthaler}, {Morley}, {Morris}, {Mulone}, {Muraveva}, {Musella},
  {Narbonne}, {Nelemans}, {Nicastro}, {Noval}, {Ord{\'e}novic},
  {Ordieres-Mer{\'e}}, {Osborne}, {Pagani}, {Pagano}, {Pailler}, {Palacin},
  {Palaversa}, {Parsons}, {Paulsen}, {Pecoraro}, {Pedrosa}, {Pentik{\"a}inen},
  {Pereira}, {Pichon}, {Piersimoni}, {Pineau}, {Plachy}, {Plum}, {Poujoulet},
  {Pr{\v{s}}a}, {Pulone}, {Ragaini}, {Rago}, {Rambaux}, {Ramos-Lerate},
  {Ranalli}, {Rauw}, {Read}, {Regibo}, {Renk}, {Reyl{\'e}}, {Ribeiro},
  {Rimoldini}, {Ripepi}, {Riva}, {Rixon}, {Roelens}, {Romero-G{\'o}mez},
  {Rowell}, {Royer}, {Rudolph}, {Ruiz-Dern}, {Sadowski}, {Sagrist{\`a}
  Sell{\'e}s}, {Sahlmann}, {Salgado}, {Salguero}, {Sarasso}, {Savietto},
  {Schnorhk}, {Schultheis}, {Sciacca}, {Segol}, {Segovia}, {Segransan},
  {Serpell}, {Shih}, {Smareglia}, {Smart}, {Smith}, {Solano}, {Solitro},
  {Sordo}, {Soria Nieto}, {Souchay}, {Spagna}, {Spoto}, {Stampa}, {Steele},
  {Steidelm{\"u}ller}, {Stephenson}, {Stoev}, {Suess}, {S{\"u}veges}, {Surdej},
  {Szabados}, {Szegedi-Elek}, {Tapiador}, {Taris}, {Tauran}, {Taylor},
  {Teixeira}, {Terrett}, {Tingley}, {Trager}, {Turon}, {Ulla}, {Utrilla},
  {Valentini}, {van Elteren}, {Van Hemelryck}, {van Leeuwen}, {Varadi},
  {Vecchiato}, {Veljanoski}, {Via}, {Vicente}, {Vogt}, {Voss}, {Votruba},
  {Voutsinas}, {Walmsley}, {Weiler}, {Weingrill}, {Werner}, {Wevers},
  {Whitehead}, {Wyrzykowski}, {Yoldas}, {{\v{Z}}erjal}, {Zucker}, {Zurbach},
  {Zwitter}, {Alecu}, {Allen}, {Allende Prieto}, {Amorim},
  {Anglada-Escud{\'e}}, {Arsenijevic}, {Azaz}, {Balm}, {Beck}, {Bernstein},
  {Bigot}, {Bijaoui}, {Blasco}, {Bonfigli}, {Bono}, {Boudreault}, {Bressan},
  {Brown}, {Brunet}, {Bunclark}, {Buonanno}, {Butkevich}, {Carret}, {Carrion},
  {Chemin}, {Ch{\'e}reau}, {Corcione}, {Darmigny}, {de Boer}, {de Teodoro}, {de
  Zeeuw}, {Delle Luche}, {Domingues}, {Dubath}, {Fodor}, {Fr{\'e}zouls},
  {Fries}, {Fustes}, {Fyfe}, {Gallardo}, {Gallegos}, {Gardiol}, {Gebran},
  {Gomboc}, {G{\'o}mez}, {Grux}, {Gueguen}, {Heyrovsky}, {Hoar}, {Iannicola},
  {Isasi Parache}, {Janotto}, {Joliet}, {Jonckheere}, {Keil}, {Kim},
  {Klagyivik}, {Klar}, {Knude}, {Kochukhov}, {Kolka}, {Kos}, {Kutka}, {Lainey},
  {LeBouquin}, {Liu}, {Loreggia}, {Makarov}, {Marseille}, {Martayan},
  {Martinez-Rubi}, {Massart}, {Meynadier}, {Mignot}, {Munari}, {Nguyen},
  {Nordlander}, {Ocvirk}, {O'Flaherty}, {Olias Sanz}, {Ortiz}, {Osorio},
  {Oszkiewicz}, {Ouzounis}, {Palmer}, {Park}, {Pasquato}, {Peltzer}, {Peralta},
  {P{\'e}turaud}, {Pieniluoma}, {Pigozzi}, {Poels}, {Prat}, {Prod'homme},
  {Raison}, {Rebordao}, {Risquez}, {Rocca-Volmerange}, {Rosen}, {Ruiz-Fuertes},
  {Russo}, {Sembay}, {Serraller Vizcaino}, {Short}, {Siebert}, {Silva},
  {Sinachopoulos}, {Slezak}, {Soffel}, {Sosnowska}, {Strai{\v{z}}ys}, {ter
  Linden}, {Terrell}, {Theil}, {Tiede}, {Troisi}, {Tsalmantza}, {Tur},
  {Vaccari}, {Vachier}, {Valles}, {Van Hamme}, {Veltz}, {Virtanen}, {Wallut},
  {Wichmann}, {Wilkinson}, {Ziaeepour}, \& {Zschocke}}]{gaia1}
{Gaia Collaboration}, {Prusti}, T., {de Bruijne}, J.~H.~J., {et~al.} 2016,
  \aap, 595, A1, \dodoi{10.1051/0004-6361/201629272}

\bibitem[{{Gaia Collaboration} {et~al.}(2021){Gaia Collaboration}, {Brown},
  {Vallenari}, {Prusti}, {de Bruijne}, {Babusiaux}, {Biermann}, {Creevey},
  {Evans}, {Eyer}, {Hutton}, {Jansen}, {Jordi}, {Klioner}, {Lammers},
  {Lindegren}, {Luri}, {Mignard}, {Panem}, {Pourbaix}, {Randich}, {Sartoretti},
  {Soubiran}, {Walton}, {Arenou}, {Bailer-Jones}, {Bastian}, {Cropper},
  {Drimmel}, {Katz}, {Lattanzi}, {van Leeuwen}, {Bakker}, {Cacciari},
  {Casta{\~n}eda}, {De Angeli}, {Ducourant}, {Fabricius}, {Fouesneau},
  {Fr{\'e}mat}, {Guerra}, {Guerrier}, {Guiraud}, {Jean-Antoine Piccolo},
  {Masana}, {Messineo}, {Mowlavi}, {Nicolas}, {Nienartowicz}, {Pailler},
  {Panuzzo}, {Riclet}, {Roux}, {Seabroke}, {Sordo}, {Tanga}, {Th{\'e}venin},
  {Gracia-Abril}, {Portell}, {Teyssier}, {Altmann}, {Andrae}, {Bellas-Velidis},
  {Benson}, {Berthier}, {Blomme}, {Brugaletta}, {Burgess}, {Busso}, {Carry},
  {Cellino}, {Cheek}, {Clementini}, {Damerdji}, {Davidson}, {Delchambre},
  {Dell'Oro}, {Fern{\'a}ndez-Hern{\'a}ndez}, {Galluccio}, {Garc{\'\i}a-Lario},
  {Garcia-Reinaldos}, {Gonz{\'a}lez-N{\'u}{\~n}ez}, {Gosset}, {Haigron},
  {Halbwachs}, {Hambly}, {Harrison}, {Hatzidimitriou}, {Heiter},
  {Hern{\'a}ndez}, {Hestroffer}, {Hodgkin}, {Holl}, {Jan{\ss}en}, {Jevardat de
  Fombelle}, {Jordan}, {Krone-Martins}, {Lanzafame}, {L{\"o}ffler}, {Lorca},
  {Manteiga}, {Marchal}, {Marrese}, {Moitinho}, {Mora}, {Muinonen}, {Osborne},
  {Pancino}, {Pauwels}, {Petit}, {Recio-Blanco}, {Richards}, {Riello},
  {Rimoldini}, {Robin}, {Roegiers}, {Rybizki}, {Sarro}, {Siopis}, {Smith},
  {Sozzetti}, {Ulla}, {Utrilla}, {van Leeuwen}, {van Reeven}, {Abbas}, {Abreu
  Aramburu}, {Accart}, {Aerts}, {Aguado}, {Ajaj}, {Altavilla}, {{\'A}lvarez},
  {{\'A}lvarez Cid-Fuentes}, {Alves}, {Anderson}, {Anglada Varela}, {Antoja},
  {Audard}, {Baines}, {Baker}, {Balaguer-N{\'u}{\~n}ez}, {Balbinot}, {Balog},
  {Barache}, {Barbato}, {Barros}, {Barstow}, {Bartolom{\'e}}, {Bassilana},
  {Bauchet}, {Baudesson-Stella}, {Becciani}, {Bellazzini}, {Bernet}, {Bertone},
  {Bianchi}, {Blanco-Cuaresma}, {Boch}, {Bombrun}, {Bossini}, {Bouquillon},
  {Bragaglia}, {Bramante}, {Breedt}, {Bressan}, {Brouillet}, {Bucciarelli},
  {Burlacu}, {Busonero}, {Butkevich}, {Buzzi}, {Caffau}, {Cancelliere},
  {C{\'a}novas}, {Cantat-Gaudin}, {Carballo}, {Carlucci}, {Carnerero},
  {Carrasco}, {Casamiquela}, {Castellani}, {Castro-Ginard}, {Castro Sampol},
  {Chaoul}, {Charlot}, {Chemin}, {Chiavassa}, {Cioni}, {Comoretto}, {Cooper},
  {Cornez}, {Cowell}, {Crifo}, {Crosta}, {Crowley}, {Dafonte}, {Dapergolas},
  {David}, {David}, {de Laverny}, {De Luise}, {De March}, {De Ridder}, {de
  Souza}, {de Teodoro}, {de Torres}, {del Peloso}, {del Pozo}, {Delbo},
  {Delgado}, {Delgado}, {Delisle}, {Di Matteo}, {Diakite}, {Diener},
  {Distefano}, {Dolding}, {Eappachen}, {Edvardsson}, {Enke}, {Esquej}, {Fabre},
  {Fabrizio}, {Faigler}, {Fedorets}, {Fernique}, {Fienga}, {Figueras},
  {Fouron}, {Fragkoudi}, {Fraile}, {Franke}, {Gai}, {Garabato},
  {Garcia-Gutierrez}, {Garc{\'\i}a-Torres}, {Garofalo}, {Gavras}, {Gerlach},
  {Geyer}, {Giacobbe}, {Gilmore}, {Girona}, {Giuffrida}, {Gomel}, {Gomez},
  {Gonzalez-Santamaria}, {Gonz{\'a}lez-Vidal}, {Granvik},
  {Guti{\'e}rrez-S{\'a}nchez}, {Guy}, {Hauser}, {Haywood}, {Helmi}, {Hidalgo},
  {Hilger}, {H{\l}adczuk}, {Hobbs}, {Holland}, {Huckle}, {Jasniewicz},
  {Jonker}, {Juaristi Campillo}, {Julbe}, {Karbevska}, {Kervella}, {Khanna},
  {Kochoska}, {Kontizas}, {Kordopatis}, {Korn}, {Kostrzewa-Rutkowska},
  {Kruszy{\'n}ska}, {Lambert}, {Lanza}, {Lasne}, {Le Campion}, {Le Fustec},
  {Lebreton}, {Lebzelter}, {Leccia}, {Leclerc}, {Lecoeur-Taibi}, {Liao},
  {Licata}, {Lindstr{\o}m}, {Lister}, {Livanou}, {Lobel}, {Madrero Pardo},
  {Managau}, {Mann}, {Marchant}, {Marconi}, {Marcos Santos}, {Marinoni},
  {Marocco}, {Marshall}, {Martin Polo}, {Mart{\'\i}n-Fleitas}, {Masip},
  {Massari}, {Mastrobuono-Battisti}, {Mazeh}, {McMillan}, {Messina},
  {Michalik}, {Millar}, {Mints}, {Molina}, {Molinaro}, {Moln{\'a}r},
  {Montegriffo}, {Mor}, {Morbidelli}, {Morel}, {Morris}, {Mulone}, {Munoz},
  {Muraveva}, {Murphy}, {Musella}, {Noval}, {Ord{\'e}novic}, {Orr{\`u}},
  {Osinde}, {Pagani}, {Pagano}, {Palaversa}, {Palicio}, {Panahi}, {Pawlak},
  {Pe{\~n}alosa Esteller}, {Penttil{\"a}}, {Piersimoni}, {Pineau}, {Plachy},
  {Plum}, {Poggio}, {Poretti}, {Poujoulet}, {Pr{\v{s}}a}, {Pulone}, {Racero},
  {Ragaini}, {Rainer}, {Raiteri}, {Rambaux}, {Ramos}, {Ramos-Lerate}, {Re
  Fiorentin}, {Regibo}, {Reyl{\'e}}, {Ripepi}, {Riva}, {Rixon}, {Robichon},
  {Robin}, {Roelens}, {Rohrbasser}, {Romero-G{\'o}mez}, {Rowell}, {Royer},
  {Rybicki}, {Sadowski}, {Sagrist{\`a} Sell{\'e}s}, {Sahlmann}, {Salgado},
  {Salguero}, {Samaras}, {Sanchez Gimenez}, {Sanna}, {Santove{\~n}a},
  {Sarasso}, {Schultheis}, {Sciacca}, {Segol}, {Segovia}, {S{\'e}gransan},
  {Semeux}, {Shahaf}, {Siddiqui}, {Siebert}, {Siltala}, {Slezak}, {Smart},
  {Solano}, {Solitro}, {Souami}, {Souchay}, {Spagna}, {Spoto}, {Steele},
  {Steidelm{\"u}ller}, {Stephenson}, {S{\"u}veges}, {Szabados}, {Szegedi-Elek},
  {Taris}, {Tauran}, {Taylor}, {Teixeira}, {Thuillot}, {Tonello}, {Torra},
  {Torra}, {Turon}, {Unger}, {Vaillant}, {van Dillen}, {Vanel}, {Vecchiato},
  {Viala}, {Vicente}, {Voutsinas}, {Weiler}, {Wevers}, {Wyrzykowski}, {Yoldas},
  {Yvard}, {Zhao}, {Zorec}, {Zucker}, {Zurbach}, \& {Zwitter}}]{gaia2}
{Gaia Collaboration}, {Brown}, A.~G.~A., {Vallenari}, A., {et~al.} 2021, \aap,
  649, A1, \dodoi{10.1051/0004-6361/202039657}

\bibitem[{{Galametz} {et~al.}(2018){Galametz}, {Maury}, {Girart}, {Rao},
  {Zhang}, {Gaudel}, {Valdivia}, {Keto}, \& {Lai}}]{galametz2018}
{Galametz}, M., {Maury}, A., {Girart}, J.~M., {et~al.} 2018, \aap, 616, A139,
  \dodoi{10.1051/0004-6361/201833004}

\bibitem[{{Galametz} {et~al.}(2020){Galametz}, {Maury}, {Girart}, {Rao},
  {Zhang}, {Gaudel}, {Valdivia}, {Hennebelle}, {Cabedo-Soto}, {Keto}, \&
  {Lai}}]{galametz2020}
---. 2020, \aap, 644, A47, \dodoi{10.1051/0004-6361/202038854}

\bibitem[{{Girart} {et~al.}(2006){Girart}, {Rao}, \& {Marrone}}]{girart06}
{Girart}, J.~M., {Rao}, R., \& {Marrone}, D.~P. 2006, Science, 313, 812,
  \dodoi{10.1126/science.1129093}

\bibitem[{{Goodman} {et~al.}(1995){Goodman}, {Jones}, {Lada}, \&
  {Myers}}]{goodman1995}
{Goodman}, A.~A., {Jones}, T.~J., {Lada}, E.~A., \& {Myers}, P.~C. 1995, \apj,
  448, 748, \dodoi{10.1086/176003}

\bibitem[{{Gordon} {et~al.}(2018){Gordon}, {Lopez-Rodriguez}, {Andersson},
  {Clarke}, {Coude}, {Moullet}, {Richards}, {Shuping}, {Vacca}, \&
  {Yorke}}]{gordon2018}
{Gordon}, M.~S., {Lopez-Rodriguez}, E., {Andersson}, B.~G., {et~al.} 2018,
  arXiv e-prints, arXiv:1811.03100.
\newblock \doarXiv{1811.03100}

\bibitem[{{Harper} {et~al.}(2018){Harper}, {Runyan}, {Dowell}, {Wirth},
  {Amato}, {Ames}, {Amiri}, {Banks}, {Bartels}, {Benford}, {Berthoud},
  {Buchanan}, {Casey}, {Chapman}, {Chuss}, {Cook}, {Derro}, {Dotson}, {Evans},
  {Fixsen}, {Gatley}, {Guerra}, {Halpern}, {Hamilton}, {Hamlin}, {Hansen},
  {Heimsath}, {Hermida}, {Hilton}, {Hirsch}, {Hollister}, {Hostetter}, {Irwin},
  {Jhabvala}, {Jhabvala}, {Kastner}, {Kov{\'a}cs}, {Lin}, {Loewenstein},
  {Looney}, {Lopez-Rodriguez}, {Maher}, {Michail}, {Miller}, {Moseley},
  {Novak}, {Pernic}, {Rennick}, {Rhody}, {Sandberg}, {Sandford}, {Santos},
  {Shafer}, {Sharp}, {Shirron}, {Siah}, {Silverberg}, {Sparr}, {Spotz},
  {Staguhn}, {Toorian}, {Towey}, {Tuttle}, {Vaillancourt}, {Voellmer},
  {Volpert}, {Wang}, \& {Wollack}}]{harper2018}
{Harper}, D.~A., {Runyan}, M.~C., {Dowell}, C.~D., {et~al.} 2018, Journal of
  Astronomical Instrumentation, 7, 1840008, \dodoi{10.1142/S2251171718400081}

\bibitem[{{Hennebelle} {et~al.}(2011){Hennebelle}, {Commer{\c{c}}on}, {Joos},
  {Klessen}, {Krumholz}, {Tan}, \& {Teyssier}}]{hennebelle2011}
{Hennebelle}, P., {Commer{\c{c}}on}, B., {Joos}, M., {et~al.} 2011, \aap, 528,
  A72, \dodoi{10.1051/0004-6361/201016052}

\bibitem[{{Hildebrand}(1983)}]{hildebrand1983}
{Hildebrand}, R.~H. 1983, \qjras, 24, 267

\bibitem[{{Hildebrand} {et~al.}(2000){Hildebrand}, {Davidson}, {Dotson},
  {Dowell}, {Novak}, \& {Vaillancourt}}]{hildebrand2000}
{Hildebrand}, R.~H., {Davidson}, J.~A., {Dotson}, J.~L., {et~al.} 2000, \pasp,
  112, 1215, \dodoi{10.1086/316613}

\bibitem[{{Hull} {et~al.}(2020){Hull}, {Le Gouellec}, {Girart}, {Tobin}, \&
  {Bourke}}]{hull2020}
{Hull}, C. L.~H., {Le Gouellec}, V. J.~M., {Girart}, J.~M., {Tobin}, J.~J., \&
  {Bourke}, T.~L. 2020, \apj, 892, 152, \dodoi{10.3847/1538-4357/ab5809}

\bibitem[{{Hull} \& {Zhang}(2019)}]{hull2019}
{Hull}, C. L.~H., \& {Zhang}, Q. 2019, Frontiers in Astronomy and Space
  Sciences, 6, 3, \dodoi{10.3389/fspas.2019.00003}

\bibitem[{{Hull} {et~al.}(2014){Hull}, {Plambeck}, {Kwon}, {Bower},
  {Carpenter}, {Crutcher}, {Fiege}, {Franzmann}, {Hakobian}, {Heiles}, {Houde},
  {Hughes}, {Lamb}, {Looney}, {Marrone}, {Matthews}, {Pillai}, {Pound},
  {Rahman}, {Sandell}, {Stephens}, {Tobin}, {Vaillancourt}, {Volgenau}, \&
  {Wright}}]{hull2014}
{Hull}, C. L.~H., {Plambeck}, R.~L., {Kwon}, W., {et~al.} 2014, \apjs, 213, 13,
  \dodoi{10.1088/0067-0049/213/1/13}

\bibitem[{{Hull} {et~al.}(2017){Hull}, {Girart}, {Tychoniec}, {Rao},
  {Cort{\'e}s}, {Pokhrel}, {Zhang}, {Houde}, {Dunham}, {Kristensen}, {Lai},
  {Li}, \& {Plambeck}}]{hull2017b}
{Hull}, C. L.~H., {Girart}, J.~M., {Tychoniec}, {\L}., {et~al.} 2017, \apj,
  847, 92, \dodoi{10.3847/1538-4357/aa7fe9}

\bibitem[{Hunter(2007)}]{matplotlib}
Hunter, J.~D. 2007, Computing in Science Engineering, 9, 90,
  \dodoi{10.1109/MCSE.2007.55}

\bibitem[{{Jacobsen} {et~al.}(2019){Jacobsen}, {J{\o}rgensen}, {Di Francesco},
  {Evans}, {Choi}, \& {Lee}}]{jacobsen2019}
{Jacobsen}, S.~K., {J{\o}rgensen}, J.~K., {Di Francesco}, J., {et~al.} 2019,
  \aap, 629, A29, \dodoi{10.1051/0004-6361/201833214}

\bibitem[{{Joos} {et~al.}(2012){Joos}, {Hennebelle}, \& {Ciardi}}]{joos2012}
{Joos}, M., {Hennebelle}, P., \& {Ciardi}, A. 2012, \aap, 543, A128,
  \dodoi{10.1051/0004-6361/201118730}

\bibitem[{{J{\o}rgensen}(2004)}]{jorgensen2004}
{J{\o}rgensen}, J.~K. 2004, \aap, 424, 589, \dodoi{10.1051/0004-6361:20040247}

\bibitem[{{Kwon} {et~al.}(2019){Kwon}, {Stephens}, {Tobin}, {Looney}, {Li},
  {van der Tak}, \& {Crutcher}}]{kwon2019}
{Kwon}, W., {Stephens}, I.~W., {Tobin}, J.~J., {et~al.} 2019, \apj, 879, 25,
  \dodoi{10.3847/1538-4357/ab24c8}

\bibitem[{{Launhardt} {et~al.}(2013){Launhardt}, {Stutz}, {Schmiedeke},
  {Henning}, {Krause}, {Balog}, {Beuther}, {Birkmann}, {Hennemann},
  {Kainulainen}, {Khanzadyan}, {Linz}, {Lippok}, {Nielbock}, {Pitann}, {Ragan},
  {Risacher}, {Schmalzl}, {Shirley}, {Stecklum}, {Steinacker}, \&
  {Tackenberg}}]{launhardt2013}
{Launhardt}, R., {Stutz}, A.~M., {Schmiedeke}, A., {et~al.} 2013, \aap, 551,
  A98, \dodoi{10.1051/0004-6361/201220477}

\bibitem[{{Lazarian}(2007)}]{lazarian2007}
{Lazarian}, A. 2007, \jqsrt, 106, 225, \dodoi{10.1016/j.jqsrt.2007.01.038}

\bibitem[{{Lee} {et~al.}(2021){Lee}, {Berthoud}, {Chen}, {Cox}, {Davidson},
  {Encalada}, {Fissel}, {Harrison}, {Kwon}, {Li}, {Li}, {Looney}, {Novak},
  {Sadavoy}, {Santos}, {Segura-Cox}, \& {Stephens}}]{lee2021}
{Lee}, D., {Berthoud}, M., {Chen}, C.-Y., {et~al.} 2021, \apj, 918, 39,
  \dodoi{10.3847/1538-4357/ac0cf2}

\bibitem[{{Leung} {et~al.}(2016){Leung}, {Lim}, \& {Takakuwa}}]{leung2016}
{Leung}, G. Y.~C., {Lim}, J., \& {Takakuwa}, S. 2016, \apj, 833, 55,
  \dodoi{10.3847/1538-4357/833/1/55}

\bibitem[{{Li} {et~al.}(2014{\natexlab{a}}){Li}, {Banerjee}, {Pudritz},
  {J{\o}rgensen}, {Shang}, {Krasnopolsky}, \& {Maury}}]{li2014b}
{Li}, Z.~Y., {Banerjee}, R., {Pudritz}, R.~E., {et~al.} 2014{\natexlab{a}}, in
  Protostars and Planets VI, ed. H.~{Beuther}, R.~S. {Klessen}, C.~P.
  {Dullemond}, \& T.~{Henning}, 173,
  \dodoi{10.2458/azu\_uapress\_9780816531240-ch008}

\bibitem[{{Li} {et~al.}(2011){Li}, {Krasnopolsky}, \& {Shang}}]{li2011}
{Li}, Z.-Y., {Krasnopolsky}, R., \& {Shang}, H. 2011, \apj, 738, 180,
  \dodoi{10.1088/0004-637X/738/2/180}

\bibitem[{{Li} {et~al.}(2014{\natexlab{b}}){Li}, {Krasnopolsky}, {Shang}, \&
  {Zhao}}]{li2014}
{Li}, Z.-Y., {Krasnopolsky}, R., {Shang}, H., \& {Zhao}, B. 2014{\natexlab{b}},
  \apj, 793, 130, \dodoi{10.1088/0004-637X/793/2/130}

\bibitem[{{Lindegren} {et~al.}(2021){Lindegren}, {Klioner}, {Hern{\'a}ndez},
  {Bombrun}, {Ramos-Lerate}, {Steidelm{\"u}ller}, {Bastian}, {Biermann}, {de
  Torres}, {Gerlach}, {Geyer}, {Hilger}, {Hobbs}, {Lammers}, {McMillan},
  {Stephenson}, {Casta{\~n}eda}, {Davidson}, {Fabricius}, {Gracia-Abril},
  {Portell}, {Rowell}, {Teyssier}, {Torra}, {Bartolom{\'e}}, {Clotet},
  {Garralda}, {Gonz{\'a}lez-Vidal}, {Torra}, {Abbas}, {Altmann}, {Anglada
  Varela}, {Balaguer-N{\'u}{\~n}ez}, {Balog}, {Barache}, {Becciani}, {Bernet},
  {Bertone}, {Bianchi}, {Bouquillon}, {Brown}, {Bucciarelli}, {Busonero},
  {Butkevich}, {Buzzi}, {Cancelliere}, {Carlucci}, {Charlot}, {Cioni},
  {Crosta}, {Crowley}, {del Peloso}, {del Pozo}, {Drimmel}, {Esquej}, {Fienga},
  {Fraile}, {Gai}, {Garcia-Reinaldos}, {Guerra}, {Hambly}, {Hauser},
  {Jan{\ss}en}, {Jordan}, {Kostrzewa-Rutkowska}, {Lattanzi}, {Liao}, {Licata},
  {Lister}, {L{\"o}ffler}, {Marchant}, {Masip}, {Mignard}, {Mints}, {Molina},
  {Mora}, {Morbidelli}, {Murphy}, {Pagani}, {Panuzzo}, {Pe{\~n}alosa Esteller},
  {Poggio}, {Re Fiorentin}, {Riva}, {Sagrist{\`a} Sell{\'e}s}, {Sanchez
  Gimenez}, {Sarasso}, {Sciacca}, {Siddiqui}, {Smart}, {Souami}, {Spagna},
  {Steele}, {Taris}, {Utrilla}, {van Reeven}, \& {Vecchiato}}]{gaia3}
{Lindegren}, L., {Klioner}, S.~A., {Hern{\'a}ndez}, J., {et~al.} 2021, \aap,
  649, A2, \dodoi{10.1051/0004-6361/202039709}

\bibitem[{{Looney} {et~al.}(2007){Looney}, {Tobin}, \& {Kwon}}]{looney2007}
{Looney}, L.~W., {Tobin}, J.~J., \& {Kwon}, W. 2007, \apjl, 670, L131,
  \dodoi{10.1086/524361}

\bibitem[{{Lynds}(1962)}]{lynds1962}
{Lynds}, B.~T. 1962, \apjs, 7, 1, \dodoi{10.1086/190072}

\bibitem[{{Magalhaes} {et~al.}(1996){Magalhaes}, {Rodrigues}, {Margoniner},
  {Pereyra}, \& {Heathcote}}]{Magalhaes1996}
{Magalhaes}, A.~M., {Rodrigues}, C.~V., {Margoniner}, V.~E., {Pereyra}, A., \&
  {Heathcote}, S. 1996, in Astronomical Society of the Pacific Conference
  Series, Vol.~97, Polarimetry of the Interstellar Medium, ed. W.~G. {Roberge}
  \& D.~C.~B. {Whittet}, 118

\bibitem[{{Maury} {et~al.}(2018){Maury}, {Girart}, {Zhang}, {Hennebelle},
  {Keto}, {Rao}, {Lai}, {Ohashi}, \& {Galametz}}]{maury2018}
{Maury}, A.~J., {Girart}, J.~M., {Zhang}, Q., {et~al.} 2018, \mnras, 477, 2760,
  \dodoi{10.1093/mnras/sty574}

\bibitem[{{McKee} \& {Ostriker}(2007)}]{mckee2007}
{McKee}, C.~F., \& {Ostriker}, E.~C. 2007, \araa, 45, 565,
  \dodoi{10.1146/annurev.astro.45.051806.110602}

\bibitem[{{McMullin} {et~al.}(2007){McMullin}, {Waters}, {Schiebel}, {Young},
  \& {Golap}}]{casa}
{McMullin}, J.~P., {Waters}, B., {Schiebel}, D., {Young}, W., \& {Golap}, K.
  2007, in Astronomical Society of the Pacific Conference Series, Vol. 376,
  Astronomical Data Analysis Software and Systems XVI, ed. R.~A. {Shaw},
  F.~{Hill}, \& D.~J. {Bell}, 127

\bibitem[{{Mignon-Risse} {et~al.}(2021){Mignon-Risse}, {Gonz{\'a}lez}, \&
  {Commer{\c{c}}on}}]{mignonrisse2021}
{Mignon-Risse}, R., {Gonz{\'a}lez}, M., \& {Commer{\c{c}}on}, B. 2021, arXiv
  e-prints, arXiv:2109.11241.
\newblock \doarXiv{2109.11241}

\bibitem[{{Mocz} {et~al.}(2017){Mocz}, {Burkhart}, {Hernquist}, {McKee}, \&
  {Springel}}]{mocz2017}
{Mocz}, P., {Burkhart}, B., {Hernquist}, L., {McKee}, C.~F., \& {Springel}, V.
  2017, \apj, 838, 40, \dodoi{10.3847/1538-4357/aa6475}

\bibitem[{{Monsch} {et~al.}(2018){Monsch}, {Pineda}, {Liu}, {Zucker}, {How-Huan
  Chen}, {Pattle}, {Offner}, {Di Francesco}, {Ginsburg}, {Ercolano}, {Arce},
  {Friesen}, {Kirk}, {Caselli}, \& {Goodman}}]{monsch2018}
{Monsch}, K., {Pineda}, J.~E., {Liu}, H.~B., {et~al.} 2018, \apj, 861, 77,
  \dodoi{10.3847/1538-4357/aac8da}

\bibitem[{{Mouschovias} \& {Paleologou}(1979)}]{mouschovias1979}
{Mouschovias}, T.~C., \& {Paleologou}, E.~V. 1979, \apj, 230, 204,
  \dodoi{10.1086/157077}

\bibitem[{{Myers} {et~al.}(2020){Myers}, {Stephens}, {Auddy}, {Basu}, {Bourke},
  \& {Hull}}]{myers2020}
{Myers}, P.~C., {Stephens}, I.~W., {Auddy}, S., {et~al.} 2020, \apj, 896, 163,
  \dodoi{10.3847/1538-4357/ab9110}

\bibitem[{{NHSC}(2020{\natexlab{a}})}]{herschel250doi}
{NHSC}. 2020{\natexlab{a}}, Herschel SPIRE Point Source Catalog: 250 microns,
  IPAC, \dodoi{10.26131/IRSA45}

\bibitem[{{NHSC}(2020{\natexlab{b}})}]{herschel350doi}
---. 2020{\natexlab{b}}, Herschel SPIRE Point Source Catalog: 350 microns,
  IPAC, \dodoi{10.26131/IRSA46}

\bibitem[{{NHSC}(2020{\natexlab{c}})}]{herschel500doi}
---. 2020{\natexlab{c}}, Herschel SPIRE Point Source Catalog: 500 microns,
  IPAC, \dodoi{10.26131/IRSA44}

\bibitem[{{Novak}(2011)}]{novak2011}
{Novak}, G. 2011, in Astronomical Society of the Pacific Conference Series,
  Vol. 449, Astronomical Polarimetry 2008: Science from Small to Large
  Telescopes, ed. P.~{Bastien}, N.~{Manset}, D.~P. {Clemens}, \& N.~{St-Louis},
  50

\bibitem[{{Offner} {et~al.}(2016){Offner}, {Dunham}, {Lee}, {Arce}, \&
  {Fielding}}]{offner2016}
{Offner}, S. S.~R., {Dunham}, M.~M., {Lee}, K.~I., {Arce}, H.~G., \&
  {Fielding}, D.~B. 2016, \apjl, 827, L11, \dodoi{10.3847/2041-8205/827/1/L11}

\bibitem[{{Ortiz-Le{\'o}n} {et~al.}(2018){Ortiz-Le{\'o}n}, {Loinard}, {Dzib},
  {Kounkel}, {Galli}, {Tobin}, {Evans}, {Hartmann}, {Rodr{\'\i}guez},
  {Brice{\~n}o}, {Torres}, \& {Mioduszewski}}]{ortiz2018}
{Ortiz-Le{\'o}n}, G.~N., {Loinard}, L., {Dzib}, S.~A., {et~al.} 2018, \apjl,
  869, L33, \dodoi{10.3847/2041-8213/aaf6ad}

\bibitem[{{Oya} {et~al.}(2018){Oya}, {Sakai}, {Watanabe}, {L{\'o}pez-Sepulcre},
  {Ceccarelli}, {Lefloch}, \& {Yamamoto}}]{oya2018}
{Oya}, Y., {Sakai}, N., {Watanabe}, Y., {et~al.} 2018, \apj, 863, 72,
  \dodoi{10.3847/1538-4357/aacf42}

\bibitem[{{Oya} {et~al.}(2017){Oya}, {Sakai}, {Watanabe}, {Higuchi}, {Hirota},
  {L{\'o}pez-Sepulcre}, {Sakai}, {Aikawa}, {Ceccarelli}, {Lefloch}, {Caux},
  {Vastel}, {Kahane}, \& {Yamamoto}}]{oya2017}
---. 2017, \apj, 837, 174, \dodoi{10.3847/1538-4357/aa6300}

\bibitem[{{Park} {et~al.}(2000){Park}, {Panis}, {Ohashi}, {Choi}, \&
  {Minh}}]{park2000}
{Park}, Y.~S., {Panis}, J.~F., {Ohashi}, N., {Choi}, M., \& {Minh}, Y.~C. 2000,
  \apj, 542, 344, \dodoi{10.1086/309501}

\bibitem[{{Parker}(1988)}]{parker1988}
{Parker}, N.~D. 1988, \mnras, 235, 139, \dodoi{10.1093/mnras/235.1.139}

\bibitem[{{Pattle} {et~al.}(2021){Pattle}, {Lai}, {Di Francesco}, {Sadavoy},
  {Ward-Thompson}, {Johnstone}, {Hoang}, {Arzoumanian}, {Bastien}, {Bourke},
  {Coud{\'e}}, {Doi}, {Eswaraiah}, {Fanciullo}, {Furuya}, {Hwang}, {Hull},
  {Kang}, {Kim}, {Kirchschlager}, {Kwon}, {Kwon}, {Lee}, {Liu}, {Redman},
  {Soam}, {Tahani}, {Tamura}, \& {Tang}}]{pattle2021}
{Pattle}, K., {Lai}, S.-P., {Di Francesco}, J., {et~al.} 2021, \apj, 907, 88,
  \dodoi{10.3847/1538-4357/abcc6c}

\bibitem[{{Pillai} {et~al.}(2020){Pillai}, {Clemens}, {Reissl}, {Myers},
  {Kauffmann}, {Lopez-Rodriguez}, {Alves}, {Franco}, {Henshaw}, {Menten},
  {Nakamura}, {Seifried}, {Sugitani}, \& {Wiesemeyer}}]{pillai2020}
{Pillai}, T. G.~S., {Clemens}, D.~P., {Reissl}, S., {et~al.} 2020, Nature
  Astronomy, 4, 1195, \dodoi{10.1038/s41550-020-1172-6}

\bibitem[{{Planck Collaboration} {et~al.}(2014){Planck Collaboration},
  {Abergel}, {Ade}, {Aghanim}, {Alves}, {Aniano}, {Armitage-Caplan}, {Arnaud},
  {Ashdown}, {Atrio-Barandela}, {Aumont}, {Baccigalupi}, {Banday}, {Barreiro},
  {Bartlett}, {Battaner}, {Benabed}, {Beno{\^\i}t}, {Benoit-L{\'e}vy},
  {Bernard}, {Bersanelli}, {Bielewicz}, {Bobin}, {Bock}, {Bonaldi}, {Bond},
  {Borrill}, {Bouchet}, {Boulanger}, {Bridges}, {Bucher}, {Burigana}, {Butler},
  {Cardoso}, {Catalano}, {Chamballu}, {Chary}, {Chiang}, {Chiang},
  {Christensen}, {Church}, {Clemens}, {Clements}, {Colombi}, {Colombo},
  {Combet}, {Couchot}, {Coulais}, {Crill}, {Curto}, {Cuttaia}, {Danese},
  {Davies}, {Davis}, {de Bernardis}, {de Rosa}, {de Zotti}, {Delabrouille},
  {Delouis}, {D{\'e}sert}, {Dickinson}, {Diego}, {Dole}, {Donzelli},
  {Dor{\'e}}, {Douspis}, {Draine}, {Dupac}, {Efstathiou}, {En{\ss}lin},
  {Eriksen}, {Falgarone}, {Finelli}, {Forni}, {Frailis}, {Fraisse},
  {Franceschi}, {Galeotta}, {Ganga}, {Ghosh}, {Giard}, {Giardino},
  {Giraud-H{\'e}raud}, {Gonz{\'a}lez-Nuevo}, {G{\'o}rski}, {Gratton},
  {Gregorio}, {Grenier}, {Gruppuso}, {Guillet}, {Hansen}, {Hanson}, {Harrison},
  {Helou}, {Henrot-Versill{\'e}}, {Hern{\'a}ndez-Monteagudo}, {Herranz},
  {Hildebrandt}, {Hivon}, {Hobson}, {Holmes}, {Hornstrup}, {Hovest},
  {Huffenberger}, {Jaffe}, {Jaffe}, {Jewell}, {Joncas}, {Jones}, {Juvela},
  {Keih{\"a}nen}, {Keskitalo}, {Kisner}, {Knoche}, {Knox}, {Kunz},
  {Kurki-Suonio}, {Lagache}, {L{\"a}hteenm{\"a}ki}, {Lamarre}, {Lasenby},
  {Laureijs}, {Lawrence}, {Leonardi}, {Le{\'o}n-Tavares}, {Lesgourgues},
  {Levrier}, {Liguori}, {Lilje}, {Linden-V{\o}rnle}, {L{\'o}pez-Caniego},
  {Lubin}, {Mac{\'\i}as-P{\'e}rez}, {Maffei}, {Maino}, {Mandolesi}, {Maris},
  {Marshall}, {Martin}, {Mart{\'\i}nez-Gonz{\'a}lez}, {Masi}, {Massardi},
  {Matarrese}, {Matthai}, {Mazzotta}, {McGehee}, {Melchiorri}, {Mendes},
  {Mennella}, {Migliaccio}, {Mitra}, {Miville-Desch{\^e}nes}, {Moneti},
  {Montier}, {Morgante}, {Mortlock}, {Munshi}, {Murphy}, {Naselsky}, {Nati},
  {Natoli}, {Netterfield}, {N{\o}rgaard-Nielsen}, {Noviello}, {Novikov},
  {Novikov}, {Osborne}, {Oxborrow}, {Paci}, {Pagano}, {Pajot}, {Paladini},
  {Paoletti}, {Pasian}, {Patanchon}, {Perdereau}, {Perotto}, {Perrotta},
  {Piacentini}, {Piat}, {Pierpaoli}, {Pietrobon}, {Plaszczynski},
  {Pointecouteau}, {Polenta}, {Ponthieu}, {Popa}, {Poutanen}, {Pratt},
  {Pr{\'e}zeau}, {Prunet}, {Puget}, {Rachen}, {Reach}, {Rebolo}, {Reinecke},
  {Remazeilles}, {Renault}, {Ricciardi}, {Riller}, {Ristorcelli}, {Rocha},
  {Rosset}, {Roudier}, {Rowan-Robinson}, {Rubi{\~n}o-Mart{\'\i}n}, {Rusholme},
  {Sandri}, {Santos}, {Savini}, {Scott}, {Seiffert}, {Shellard}, {Spencer},
  {Starck}, {Stolyarov}, {Stompor}, {Sudiwala}, {Sunyaev}, {Sureau}, {Sutton},
  {Suur-Uski}, {Sygnet}, {Tauber}, {Tavagnacco}, {Terenzi}, {Toffolatti},
  {Tomasi}, {Tristram}, {Tucci}, {Tuovinen}, {T{\"u}rler}, {Umana},
  {Valenziano}, {Valiviita}, {Van Tent}, {Verstraete}, {Vielva}, {Villa},
  {Vittorio}, {Wade}, {Wandelt}, {Welikala}, {Ysard}, {Yvon}, {Zacchei}, \&
  {Zonca}}]{planck2014}
{Planck Collaboration}, {Abergel}, A., {Ade}, P.~A.~R., {et~al.} 2014, \aap,
  571, A11, \dodoi{10.1051/0004-6361/201323195}

\bibitem[{{Planck Collaboration} {et~al.}(2016){Planck Collaboration}, {Ade},
  {Aghanim}, {Alves}, {Arnaud}, {Arzoumanian}, {Ashdown}, {Aumont},
  {Baccigalupi}, {Banday}, {Barreiro}, {Bartolo}, {Battaner}, {Benabed},
  {Beno{\^\i}t}, {Benoit-L{\'e}vy}, {Bernard}, {Bersanelli}, {Bielewicz},
  {Bock}, {Bonavera}, {Bond}, {Borrill}, {Bouchet}, {Boulanger}, {Bracco},
  {Burigana}, {Calabrese}, {Cardoso}, {Catalano}, {Chiang}, {Christensen},
  {Colombo}, {Combet}, {Couchot}, {Crill}, {Curto}, {Cuttaia}, {Danese},
  {Davies}, {Davis}, {de Bernardis}, {de Rosa}, {de Zotti}, {Delabrouille},
  {Dickinson}, {Diego}, {Dole}, {Donzelli}, {Dor{\'e}}, {Douspis}, {Ducout},
  {Dupac}, {Efstathiou}, {Elsner}, {En{\ss}lin}, {Eriksen},
  {Falceta-Gon{\c{c}}alves}, {Falgarone}, {Ferri{\`e}re}, {Finelli}, {Forni},
  {Frailis}, {Fraisse}, {Franceschi}, {Frejsel}, {Galeotta}, {Galli}, {Ganga},
  {Ghosh}, {Giard}, {Gjerl{\o}w}, {Gonz{\'a}lez-Nuevo}, {G{\'o}rski},
  {Gregorio}, {Gruppuso}, {Gudmundsson}, {Guillet}, {Harrison}, {Helou},
  {Hennebelle}, {Henrot-Versill{\'e}}, {Hern{\'a}ndez-Monteagudo}, {Herranz},
  {Hildebrandt}, {Hivon}, {Holmes}, {Hornstrup}, {Huffenberger}, {Hurier},
  {Jaffe}, {Jaffe}, {Jones}, {Juvela}, {Keih{\"a}nen}, {Keskitalo}, {Kisner},
  {Knoche}, {Kunz}, {Kurki-Suonio}, {Lagache}, {Lamarre}, {Lasenby},
  {Lattanzi}, {Lawrence}, {Leonardi}, {Levrier}, {Liguori}, {Lilje},
  {Linden-V{\o}rnle}, {L{\'o}pez-Caniego}, {Lubin}, {Mac{\'\i}as-P{\'e}rez},
  {Maino}, {Mandolesi}, {Mangilli}, {Maris}, {Martin},
  {Mart{\'\i}nez-Gonz{\'a}lez}, {Masi}, {Matarrese}, {Melchiorri}, {Mendes},
  {Mennella}, {Migliaccio}, {Miville-Desch{\^e}nes}, {Moneti}, {Montier},
  {Morgante}, {Mortlock}, {Munshi}, {Murphy}, {Naselsky}, {Nati},
  {Netterfield}, {Noviello}, {Novikov}, {Novikov}, {Oppermann}, {Oxborrow},
  {Pagano}, {Pajot}, {Paladini}, {Paoletti}, {Pasian}, {Perotto}, {Pettorino},
  {Piacentini}, {Piat}, {Pierpaoli}, {Pietrobon}, {Plaszczynski},
  {Pointecouteau}, {Polenta}, {Ponthieu}, {Pratt}, {Prunet}, {Puget}, {Rachen},
  {Reinecke}, {Remazeilles}, {Renault}, {Renzi}, {Ristorcelli}, {Rocha},
  {Rossetti}, {Roudier}, {Rubi{\~n}o-Mart{\'\i}n}, {Rusholme}, {Sandri},
  {Santos}, {Savelainen}, {Savini}, {Scott}, {Soler}, {Stolyarov}, {Sudiwala},
  {Sutton}, {Suur-Uski}, {Sygnet}, {Tauber}, {Terenzi}, {Toffolatti}, {Tomasi},
  {Tristram}, {Tucci}, {Umana}, {Valenziano}, {Valiviita}, {Van Tent},
  {Vielva}, {Villa}, {Wade}, {Wandelt}, {Wehus}, {Ysard}, {Yvon}, \&
  {Zonca}}]{planck2016}
{Planck Collaboration}, {Ade}, P.~A.~R., {Aghanim}, N., {et~al.} 2016, \aap,
  586, A138, \dodoi{10.1051/0004-6361/201525896}

\bibitem[{{Planck Team}(2020)}]{planck850doi}
{Planck Team}. 2020, Planck PCCS2E 353GHz Catalog,  IPAC,
  \dodoi{10.26131/IRSA461}

\bibitem[{{Pudritz} \& {Norman}(1983)}]{pudritz1983}
{Pudritz}, R.~E., \& {Norman}, C.~A. 1983, \apj, 274, 677,
  \dodoi{10.1086/161481}

\bibitem[{{Rao} {et~al.}(2009){Rao}, {Girart}, {Marrone}, {Lai}, \&
  {Schnee}}]{rao09}
{Rao}, R., {Girart}, J.~M., {Marrone}, D.~P., {Lai}, S.-P., \& {Schnee}, S.
  2009, \apj, 707, 921, \dodoi{10.1088/0004-637X/707/2/921}

\bibitem[{{Robitaille} \& {Bressert}(2012)}]{aplpy}
{Robitaille}, T., \& {Bressert}, E. 2012, {APLpy: Astronomical Plotting Library
  in Python}.
\newblock \doeprint{1208.017}

\bibitem[{{Sadavoy} {et~al.}(2013){Sadavoy}, {Di Francesco}, {Johnstone},
  {Currie}, {Drabek}, {Hatchell}, {Nutter}, {Andr{\'e}}, {Arzoumanian},
  {Benedettini}, {Bernard}, {Duarte-Cabral}, {Fallscheer}, {Friesen},
  {Greaves}, {Hennemann}, {Hill}, {Jenness}, {K{\"o}nyves}, {Matthews},
  {Mottram}, {Pezzuto}, {Roy}, {Rygl}, {Schneider-Bontemps}, {Spinoglio},
  {Testi}, {Tothill}, {Ward-Thompson}, {White}, {JCMT}, \& {Herschel Gould Belt
  Survey Teams}}]{sadavoy2013}
{Sadavoy}, S.~I., {Di Francesco}, J., {Johnstone}, D., {et~al.} 2013, \apj,
  767, 126, \dodoi{10.1088/0004-637X/767/2/126}

\bibitem[{{Sadavoy} {et~al.}(2018{\natexlab{a}}){Sadavoy}, {Myers}, {Stephens},
  {Tobin}, {Commer{\c{c}}on}, {Henning}, {Looney}, {Kwon}, {Segura-Cox}, \&
  {Harris}}]{sadavoy2018b}
{Sadavoy}, S.~I., {Myers}, P.~C., {Stephens}, I.~W., {et~al.}
  2018{\natexlab{a}}, \apj, 859, 165, \dodoi{10.3847/1538-4357/aac21a}

\bibitem[{{Sadavoy} {et~al.}(2018{\natexlab{b}}){Sadavoy}, {Myers}, {Stephens},
  {Tobin}, {Kwon}, {Segura-Cox}, {Henning}, {Commer{\c{c}}on}, \&
  {Looney}}]{sadavoy2018c}
---. 2018{\natexlab{b}}, \apj, 869, 115, \dodoi{10.3847/1538-4357/aaef81}

\bibitem[{{Sadavoy} {et~al.}(2018{\natexlab{c}}){Sadavoy}, {Keto}, {Bourke},
  {Dunham}, {Myers}, {Stephens}, {Di Francesco}, {Webb}, {Stutz}, {Launhardt},
  \& {Tobin}}]{sadavoy2018}
{Sadavoy}, S.~I., {Keto}, E., {Bourke}, T.~L., {et~al.} 2018{\natexlab{c}},
  \apj, 852, 102, \dodoi{10.3847/1538-4357/aaa080}

\bibitem[{{Santos} {et~al.}(2016){Santos}, {Busquet}, {Franco}, {Girart}, \&
  {Zhang}}]{Santos2016}
{Santos}, F.~P., {Busquet}, G., {Franco}, G. A.~P., {Girart}, J.~M., \&
  {Zhang}, Q. 2016, \apj, 832, 186, \dodoi{10.3847/0004-637X/832/2/186}

\bibitem[{{Santos} {et~al.}(2014){Santos}, {Franco}, {Roman-Lopes}, {Reis}, \&
  {Rom{\'a}n-Z{\'u}{\~n}iga}}]{Santos2014}
{Santos}, F.~P., {Franco}, G. A.~P., {Roman-Lopes}, A., {Reis}, W., \&
  {Rom{\'a}n-Z{\'u}{\~n}iga}, C.~G. 2014, \apj, 783, 1,
  \dodoi{10.1088/0004-637X/783/1/1}

\bibitem[{{Santos} {et~al.}(2012){Santos}, {Roman-Lopes}, \&
  {Franco}}]{Santos2012}
{Santos}, F.~P., {Roman-Lopes}, A., \& {Franco}, G. A.~P. 2012, \apj, 751, 138,
  \dodoi{10.1088/0004-637X/751/2/138}

\bibitem[{{Santos} {et~al.}(2017){Santos}, {Ade}, {Angil{\`e}}, {Ashton},
  {Benton}, {Devlin}, {Dober}, {Fissel}, {Fukui}, {Galitzki}, {Gandilo},
  {Klein}, {Korotkov}, {Li}, {Martin}, {Matthews}, {Moncelsi}, {Nakamura},
  {Netterfield}, {Novak}, {Pascale}, {Poidevin}, {Savini}, {Scott}, {Shariff},
  {Diego Soler}, {Thomas}, {Tucker}, {Tucker}, \& {Ward-Thompson}}]{Santos2017}
{Santos}, F.~P., {Ade}, P. A.~R., {Angil{\`e}}, F.~E., {et~al.} 2017, \apj,
  837, 161, \dodoi{10.3847/1538-4357/aa62a7}

\bibitem[{{Santos} {et~al.}(2019){Santos}, {Chuss}, {Dowell}, {Houde},
  {Looney}, {Lopez Rodriguez}, {Novak}, {Ward-Thompson}, {Berthoud}, {Dale},
  {Guerra}, {Hamilton}, {Hanany}, {Harper}, {Henning}, {Jones}, {Lazarian},
  {Michail}, {Morris}, {Staguhn}, {Stephens}, {Tassis}, {Trinh}, {Van Camp},
  {Volpert}, \& {Wollack}}]{santos2019}
{Santos}, F.~P., {Chuss}, D.~T., {Dowell}, C.~D., {et~al.} 2019, \apj, 882,
  113, \dodoi{10.3847/1538-4357/ab3407}

\bibitem[{{Sharma} {et~al.}(2020){Sharma}, {Gopinathan}, {Soam}, {Lee}, {Kim},
  {Ghosh}, {Tej}, {Kim}, {Sharma}, \& {Saha}}]{sharma2020}
{Sharma}, E., {Gopinathan}, M., {Soam}, A., {et~al.} 2020, \aap, 639, A133,
  \dodoi{10.1051/0004-6361/202037438}

\bibitem[{{Shirley} {et~al.}(2000){Shirley}, {Evans}, {Rawlings}, \&
  {Gregersen}}]{shirley2000}
{Shirley}, Y.~L., {Evans}, Neal~J., I., {Rawlings}, J. M.~C., \& {Gregersen},
  E.~M. 2000, \apjs, 131, 249, \dodoi{10.1086/317358}

\bibitem[{{Skrutskie} {et~al.}(2006){Skrutskie}, {Cutri}, {Stiening},
  {Weinberg}, {Schneider}, {Carpenter}, {Beichman}, {Capps}, {Chester},
  {Elias}, {Huchra}, {Liebert}, {Lonsdale}, {Monet}, {Price}, {Seitzer},
  {Jarrett}, {Kirkpatrick}, {Gizis}, {Howard}, {Evans}, {Fowler}, {Fullmer},
  {Hurt}, {Light}, {Kopan}, {Marsh}, {McCallon}, {Tam}, {Van Dyk}, \&
  {Wheelock}}]{2masspaper}
{Skrutskie}, M.~F., {Cutri}, R.~M., {Stiening}, R., {et~al.} 2006, \aj, 131,
  1163, \dodoi{10.1086/498708}

\bibitem[{{Soler} {et~al.}(2013){Soler}, {Hennebelle}, {Martin},
  {Miville-Desch{\^e}nes}, {Netterfield}, \& {Fissel}}]{soler2013}
{Soler}, J.~D., {Hennebelle}, P., {Martin}, P.~G., {et~al.} 2013, \apj, 774,
  128, \dodoi{10.1088/0004-637X/774/2/128}

\bibitem[{{Stephens} {et~al.}(2013){Stephens}, {Looney}, {Kwon}, {Hull},
  {Plambeck}, {Crutcher}, {Chapman}, {Novak}, {Davidson}, {Vaillancourt},
  {Shinnaga}, \& {Matthews}}]{stephens13}
{Stephens}, I.~W., {Looney}, L.~W., {Kwon}, W., {et~al.} 2013, \apjl, 769, L15,
  \dodoi{10.1088/2041-8205/769/1/L15}

\bibitem[{{Tafalla} {et~al.}(2000){Tafalla}, {Myers}, {Mardones}, \&
  {Bachiller}}]{tafalla2000}
{Tafalla}, M., {Myers}, P.~C., {Mardones}, D., \& {Bachiller}, R. 2000, \aap,
  359, 967.
\newblock \doarXiv{astro-ph/0005525}

\bibitem[{{Tobin} {et~al.}(2022){Tobin}, {Cox}, \& {Looney}}]{tobin2022}
{Tobin}, J.~J., {Cox}, E.~G., \& {Looney}, L.~W. 2022, \apj, 928, 61,
  \dodoi{10.3847/1538-4357/ac5594}

\bibitem[{{Tobin} {et~al.}(2010){Tobin}, {Hartmann}, {Looney}, \&
  {Chiang}}]{tobin2010}
{Tobin}, J.~J., {Hartmann}, L., {Looney}, L.~W., \& {Chiang}, H.-F. 2010, \apj,
  712, 1010, \dodoi{10.1088/0004-637X/712/2/1010}

\bibitem[{{Tobin} {et~al.}(2013){Tobin}, {Chandler}, {Wilner}, {Looney},
  {Loinard}, {Chiang}, {Hartmann}, {Calvet}, {D'Alessio}, {Bourke}, \&
  {Kwon}}]{tobin2013}
{Tobin}, J.~J., {Chandler}, C.~J., {Wilner}, D.~J., {et~al.} 2013, \apj, 779,
  93, \dodoi{10.1088/0004-637X/779/2/93}

\bibitem[{{Tobin} {et~al.}(2016){Tobin}, {Looney}, {Li}, {Chandler}, {Dunham},
  {Segura-Cox}, {Sadavoy}, {Melis}, {Harris}, {Kratter}, \&
  {Perez}}]{tobin2016}
{Tobin}, J.~J., {Looney}, L.~W., {Li}, Z.-Y., {et~al.} 2016, \apj, 818, 73,
  \dodoi{10.3847/0004-637X/818/1/73}

\bibitem[{van~der Walt {et~al.}(2011)van~der Walt, Colbert, \&
  Varoquaux}]{numpy}
van~der Walt, S., Colbert, S.~C., \& Varoquaux, G. 2011, Computing in Science
  Engineering, 13, 22, \dodoi{10.1109/MCSE.2011.37}

\bibitem[{{Velusamy} {et~al.}(2014){Velusamy}, {Langer}, \&
  {Thompson}}]{velusamy14}
{Velusamy}, T., {Langer}, W.~D., \& {Thompson}, T. 2014, \apj, 783, 6,
  \dodoi{10.1088/0004-637X/783/1/6}

\bibitem[{Virtanen {et~al.}(2020)Virtanen, Gommers, Oliphant, Haberland, Reddy,
  Cournapeau, Burovski, Peterson, Weckesser, Bright, {van der Walt}, Brett,
  Wilson, Millman, Mayorov, Nelson, Jones, Kern, Larson, Carey, Polat, Feng,
  Moore, {VanderPlas}, Laxalde, Perktold, Cimrman, Henriksen, Quintero, Harris,
  Archibald, Ribeiro, Pedregosa, {van Mulbregt}, \& {SciPy 1.0
  Contributors}}]{scipy}
Virtanen, P., Gommers, R., Oliphant, T.~E., {et~al.} 2020, Nature Methods, 17,
  261, \dodoi{10.1038/s41592-019-0686-2}

\bibitem[{{Wardle} \& {Kronberg}(1974)}]{Wardle1974}
{Wardle}, J.~F.~C., \& {Kronberg}, P.~P. 1974, \apj, 194, 249,
  \dodoi{10.1086/153240}

\bibitem[{{Yen} {et~al.}(2021){Yen}, {Koch}, {Hull}, {Ward-Thompson},
  {Bastien}, {Hasegawa}, {Kwon}, {Lai}, {Qiu}, {Ching}, {Chung}, {Coud{\'e}},
  {Di Francesco}, {Diep}, {Doi}, {Eswaraiah}, {Falle}, {Fuller}, {Furuya},
  {Han}, {Hatchell}, {Houde}, {Inutsuka}, {Johnstone}, {Kang}, {Kang}, {Kim},
  {Kirchschlager}, {Kwon}, {Lee}, {Lee}, {Liu}, {Liu}, {Lyo}, {Ohashi},
  {Onaka}, {Pattle}, {Sadavoy}, {Saito}, {Shinnaga}, {Soam}, {Tahani},
  {Tamura}, {Tang}, {Tang}, \& {Zhang}}]{yen2021}
{Yen}, H.-W., {Koch}, P.~M., {Hull}, C. L.~H., {et~al.} 2021, \apj, 907, 33,
  \dodoi{10.3847/1538-4357/abca99}

\bibitem[{{Zhao} {et~al.}(2018){Zhao}, {Caselli}, {Li}, \&
  {Krasnopolsky}}]{zhao2018}
{Zhao}, B., {Caselli}, P., {Li}, Z.-Y., \& {Krasnopolsky}, R. 2018, \mnras,
  473, 4868, \dodoi{10.1093/mnras/stx2617}

\bibitem[{{Zhao} \& {Li}(2013)}]{zhaoli2013}
{Zhao}, B., \& {Li}, Z.-Y. 2013, \apj, 763, 7,
  \dodoi{10.1088/0004-637X/763/1/7}

\bibitem[{{Zielinski} {et~al.}(2021){Zielinski}, {Wolf}, \&
  {Brunngr{\"a}ber}}]{zielinski2021}
{Zielinski}, N., {Wolf}, S., \& {Brunngr{\"a}ber}, R. 2021, \aap, 645, A125,
  \dodoi{10.1051/0004-6361/202039126}

\end{thebibliography}
\end{small}
\appendix 

\section{Zero Point Correction and NT fitting} \label{app:fit}
In this Appendix, we first derive zero-point corrections to the HAWC+ 154 $\mu$m and \texttt{PACS} 160 $\mu$m intensity maps. Then, we use the zero-point corrected 160 $\mu$m map together with the zero-point corrected \texttt{SPIRE} maps from the literature to derive optical depth, column density, and temperature information for L483 at 23.9$\arcsec$ resolution.

Due to the chop-nod nature of the SOFIA observations (Section \ref{sec:obs}), areas of low-level emission can become artificially low when the reference position flux is subtracted. To account for this effect, we applied a zero-point correction to our total intensity data. We created a synthetic HAWC+ Band D map by modeling the column density and temperature to the three \texttt{SPIRE} bands (as described below) and then computing from this the 154 $\mu$m flux. The \texttt{SPIRE} maps are quite a bit larger than the HAWC+ maps and cover the HAWC+ reference beam locations. We found the average reference beam flux in the synthetic 154 $\mu$m maps and added this overall flux back to our HAWC+ data, thereby creating a zero-point corrected 154 $\mu$m map.

The reference beam positions were located approximately north and south of L483. Because of this, each position in the synthetic map has a gradient of flux corresponding to which side it is on. This flux ranges within 1.74 to 2.63 mJy arcsec$^{-2}$. An average of the flux in the two reference beam positions essentially erases any gradient seen, 
and we add the mean value of 2.16 mJy arcsec$^{-2}$ to the 154 $\mu$m data. The overall value added back into the HAWC+ data accounts for $\sim$1\% of the peak flux of the source, and therefore does not significantly change any results using that value.

\texttt{SPIRE} observations were able to be zero-point corrected using the all-sky \textit{Planck} observations \citep[see][for more details]{sadavoy2018}, but the \texttt{PACS} observations were not well fit by their methods due to the small field of view compared to the \textit{Planck} beam. To account for this, we used a bootstrapping method to find the zero-point correction for our 160 $\mu$m maps. We remapped the 160 $\mu$m data onto the same pixel size as the zero-point corrected HAWC+ 154 $\mu$m data. Ignoring the 6 $\mu$m difference in wavelength, we then found the average difference in flux, off of the source, between the two observations and added it back to the \texttt{PACS} data. The added value was 19.2 MJy/sr, or 0.452 mJy arcsec$^{-2}$, and accounted for $\sim$0.1\% of the peak flux of the source at 160 $\mu$m. 

To assess whether the HAWC+ 154 $\mu$m polarization data is due to emission or absorption, we used a fitting procedure to determine the dust temperature (T) and optical depth ($\tau$) of L483. For our fit, we used the zero-point corrected \texttt{SPIRE} 250 and 350 $\mu$m \citep{sadavoy2018} maps and the zero-point corrected \texttt{PACS} 160 $\mu$m map. We prioritized the 160 $\mu$m over our 154 $\mu$m map due to its larger field-of-view. We first Gaussian-convolved and re-gridded the 160 $\mu$m and 250 $\mu$m maps to the same angular resolution as the 350 $\mu$m maps (23.9$\arcsec$). Using a fixed dust opacity spectral index \citep[$\beta = 1.62$, ][]{planck2014}, we then implemented a pixel by pixel Spectral Energy Distribution (SED) fit using a modified blackbody function,
\begin{equation}\label{eqn:sed}
    F_{\nu} = B_{\nu}(T)(1-e^{-\tau_{\nu}})\Omega,
\end{equation}
where F$_{\nu}$ is the flux in the pixel, B$_{\nu}$ is the Planck equation, $\tau_{\nu}$ is the optical depth, and $\Omega$ is the solid angle of the observations. At far-infrared wavelengths, the optical depth can be assumed to follow a power law, $\tau_{\nu}$ $\propto$ $\nu^{\beta}$. For ease, we convert Equation (\ref{eqn:sed}) to wavelength and ultimately fit,
\begin{equation}\label{eqn:sedlam}
    F_{\lambda} = \left( \frac{\lambda^{2}}{c} \right)B_{\lambda}(T)(1-e^{-\tau_{0}\left( \frac{\lambda_{0}}{\lambda} \right)^{\beta}})\Omega
\end{equation}

\begin{figure}
\includegraphics[scale = 0.4]{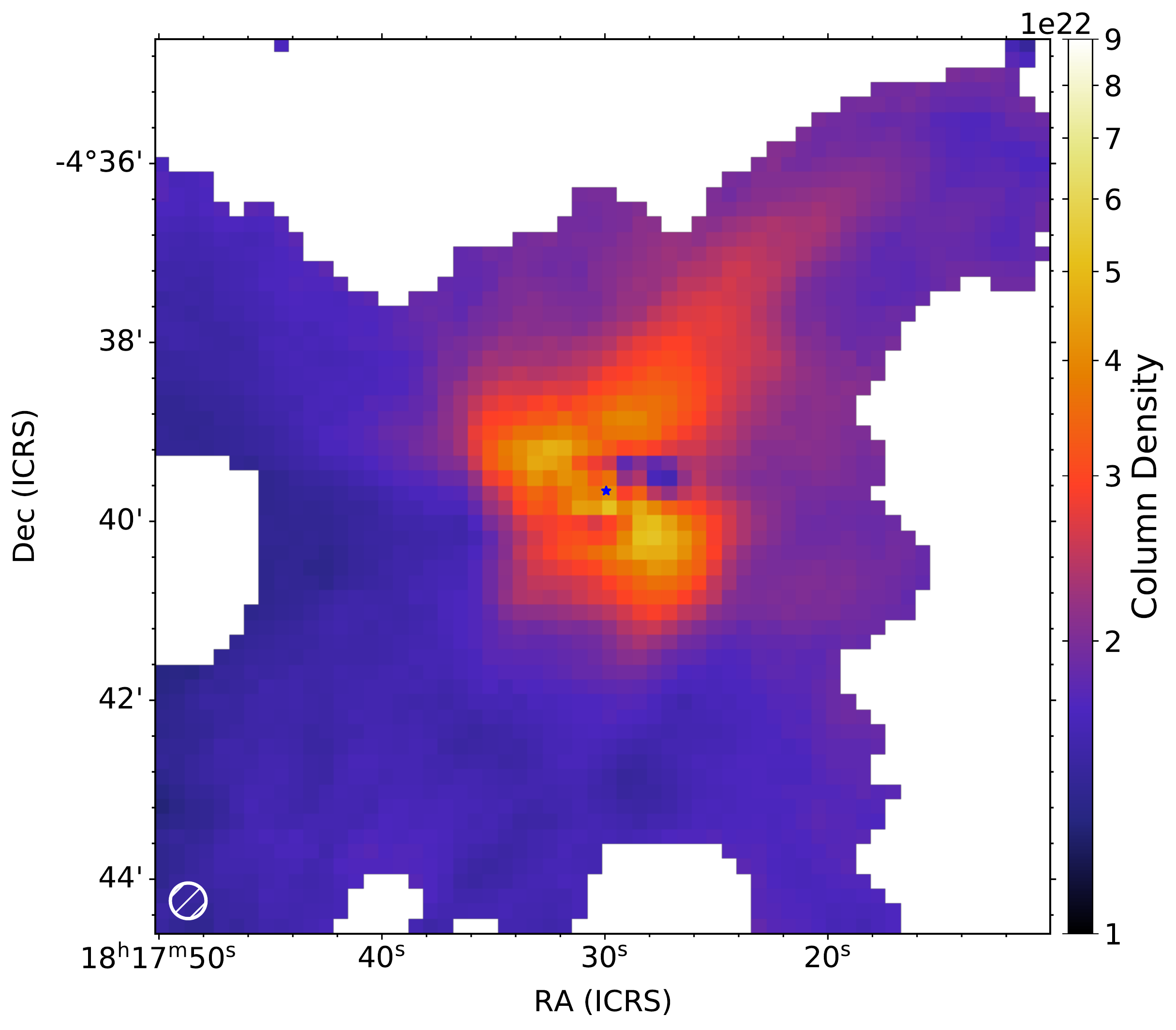}
\caption{Column Density (N) fits using \texttt{PACS} 160 $\mu$m and \texttt{SPIRE} 250 and 350 $\mu$m data shown in a log stretch to emphasis the features near the protostar. The fitting routine used a fixed $\beta$ = 1.62 value \citep{planck2014} and fit for temperature and optical depth. For ease, we assumed one temperature per sight line. To convert from optical depth ($\tau$) to N, we used an emissivity value ($\kappa$) of 0.22 cm$^{2}$g$^{-1}$, a mean molecular weight ($\mu$) of 2.8 and the mass of a Hydrogen atom in grams. The location of the ALMA protostar is shown as a blue star in the middle of the image.}
\label{fig:nh2}
\end{figure}

\begin{figure}
\includegraphics[scale = 0.4]{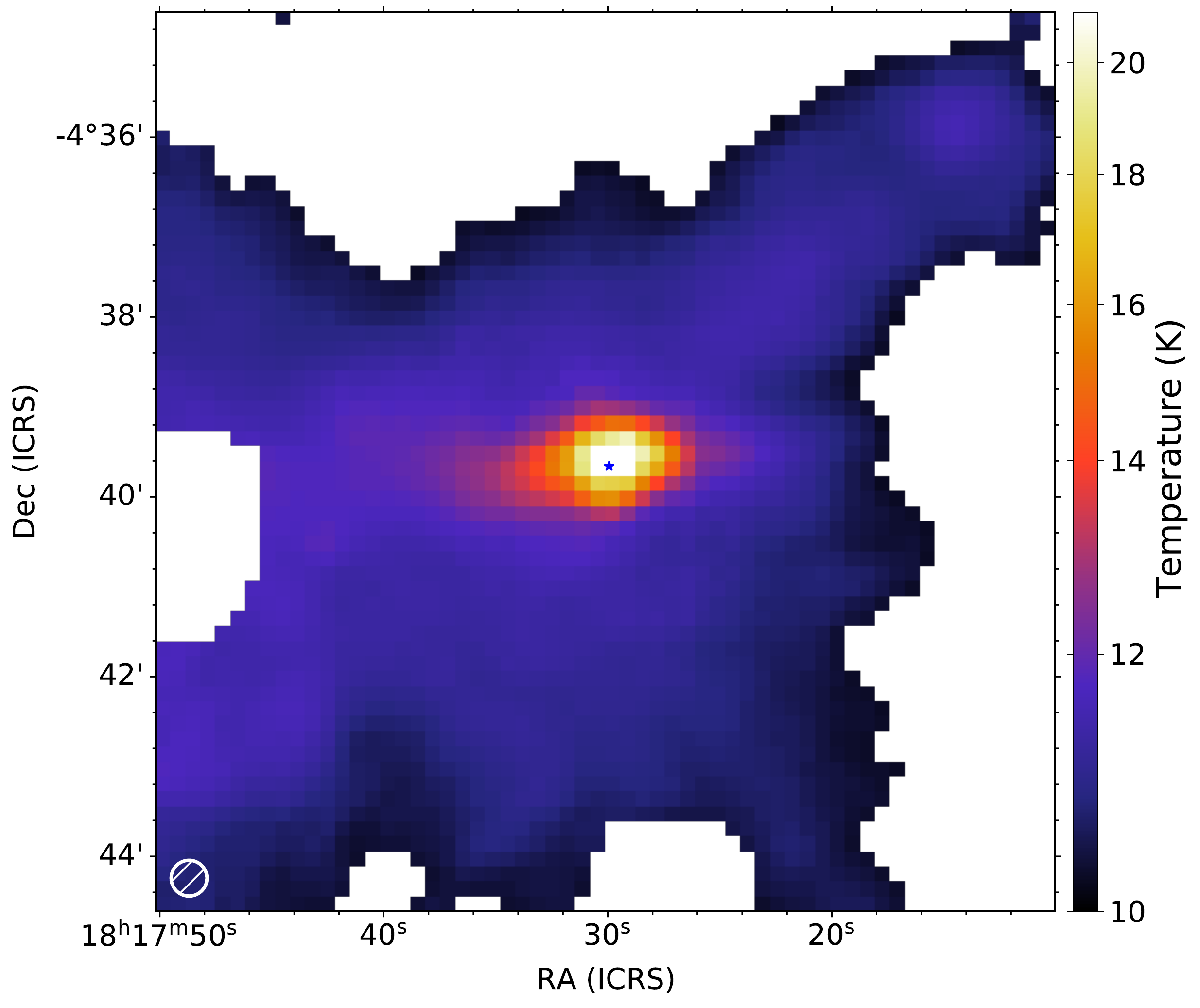}
\caption{Temperature fits using \texttt{PACS} 160 $\mu$m and \texttt{SPIRE} 250 and 350 $\mu$m data shown in a linear stretch. The fitting routine used a fixed $\beta$ = 1.62 value \citep{planck2014} and fit for temperature and optical depth. The temperature of L483 exhibits a peaked morphology and drops quickly off of the peak of the source. The location of the ALMA protostar is shown as a blue star in the middle of the image. }
\label{fig:temp}
\end{figure}

Using Equation \ref{eqn:sedlam}, we determined the dust temperature and optical depth $\tau_{0}$ at a chosen reference wavelength (i.e., $\lambda_0$ = 154 $\mu$m). To convert from optical depth to column density (N$_{H2}$), we use the relationship $\tau_{154}$ = $\kappa_{154}\mu{m}_{H}{N}_{H2}$, where $\mu$ = 2.8 is the mean molecular weight, m$_{H}$ is the mass of a hydrogen atom, and $\kappa_{154}$ is the emissivity of a dust grain at 154 $\mu$m. Using the convention for emissivity in \citet{hildebrand1983}, we find $\kappa_{154}$ = 0.22 cm$^{2}$g$^{-1}$. We account for calibration and statistical uncertainties following the method described in \citet{sadavoy2013}. In brief, for calibration uncertainty, we created 500 sets of random Gaussian distributions using a mean of 1 and a standard deviation of 0.084926 (corresponding to a half-width at half maximum of 0.1) for both \texttt{PACS} and \texttt{SPIRE} observations. To account for our statistical uncertainties, at each wavelength we generated 500 sets of random Gaussian distributions using a mean of zero and the median error associated with each band as the standard deviation. Since we use our fitted maps to determine the polarization mechanism at 154 $\mu$m and only need approximate values, we do not account for the color correction factors. Through the central pixel, we find a the column density of $\sim$ 5.2 $\times$ 10$^{22}$ cm$^{-2}$ and a temperature of $\sim$ 22 K (see Figures \ref{fig:nh2} and \ref{fig:temp}). A caveat to these results is that to truly constrain the temperature we need 850 $\mu$m data, but are unable to use \textit{Planck} results for this because its resolution (5$\arcmin$) is too large. 

The N-T fitting done by \citet{sadavoy2018} in L483 used both \texttt{SPIRE} and \texttt{PACS} data, but their results included the 500 $\mu$m band and are thus at a lower resolution. We show our N-T fits using 160, 250, and 350 $\mu$m observations in Figures \ref{fig:nh2} and \ref{fig:temp}. 
The peak temperature from \citet{sadavoy2018} is $\sim$ 18 K, which is lower than the peak temperature we find. Since the emission of L483 is more point-like than extended, we expect that this discrepancy is due to our fitting routine using higher resolution data. One trend that we do not recover in our fits is that \citet{sadavoy2018} find the temperature of L483 decreases further from the center and then increases slightly. In our column density map, we note a depletion of dust close-in to the central source, in the North-West direction. This depletion was not seen in \citet{sadavoy2018}, possibly due to the difference in angular resolution. While we cannot be sure that there are no image artifacts causing this discrepancy in morphology, we did not detect any obvious ones before or after the re-gridding process.

\section{Band E data} \label{app:bande}
Here we show the Band E (214 $\mu$m) data of L483. The peak intensity value is 118 mJy/arcsec$^{2}$, and the sensitivity in the image is 1.4 mJy/arcsec$^{2}$. This data was zero-point corrected in the same way as Band D, using a synthetic 214 $\mu$m map. The correction amount was 2.1 mJy/arcsec$^{2}$, which is $\lesssim$ 2\% of the peak value. As noted in Section \ref{sec:obs}, we were unable to use the polarization observations and thus only show the total intensity. The Band E data show a similar morphology to the Band D data, with a smearing of the source in approximately the N-S direction. As with the Band D data, we suggest this is due to the chop in the observations. While there was no significant detection of polarized emission in Band E, we report a median sensitivity of $\sigma_P$ $\sim$ 0.089 mJy arcsec$^{-2}$ .
\begin{figure}
\includegraphics[scale = 0.4]{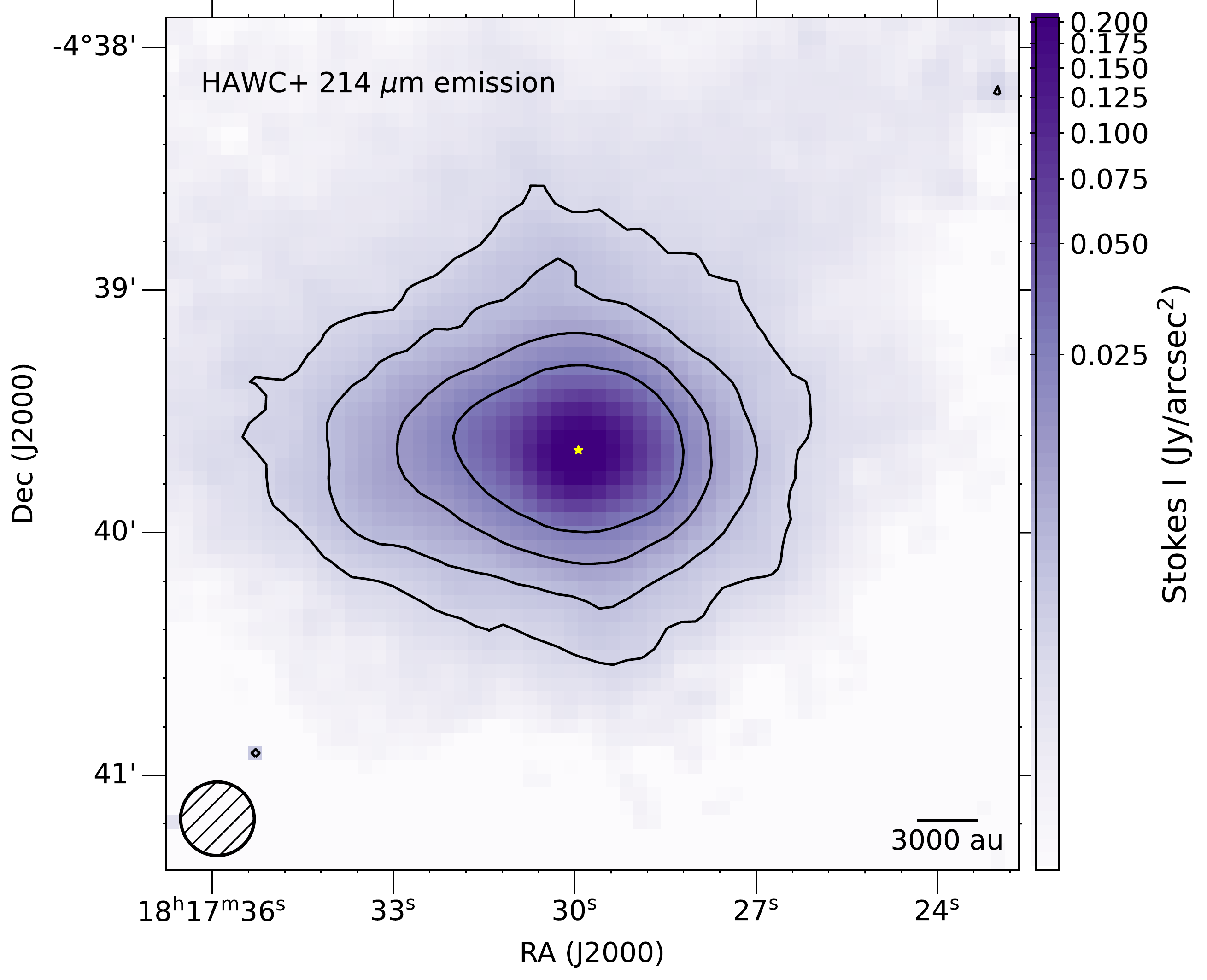}
\caption{Total intensity of 214 $\mu$m emission of L483. Contours at 3, 5, 10 and 20$\sigma$, $\sigma$ = 1.4 mJy arcsec$^{-2}$. The Band E beam (18.2 arcsec) is shown in the bottom left. The position of the ALMA protostar show as a yellow star. }
\label{fig:bandE}
\end{figure}
\section{ALMA data} \label{app:alma}
In this Appendix, we show the ALMA data using Briggs weighting with a robust parameter of 0.5. This weighting is roughly between natural (more sensitivity) and uniform (higher resolution). We performed self-calibration on this data with a solution interval of the scan time. The restoring beam of these observations is 0.137$\arcsec$ $\times$ 0.096$\arcsec$. The peak flux is 9 mJy/beam and 6 mJy/beam for the A and B protostar, respectively. Less of the material surrounding the protostars is filtered out using this image weighting, and we still easily detect the binary companion.
\begin{figure}
\includegraphics[scale = 0.4]{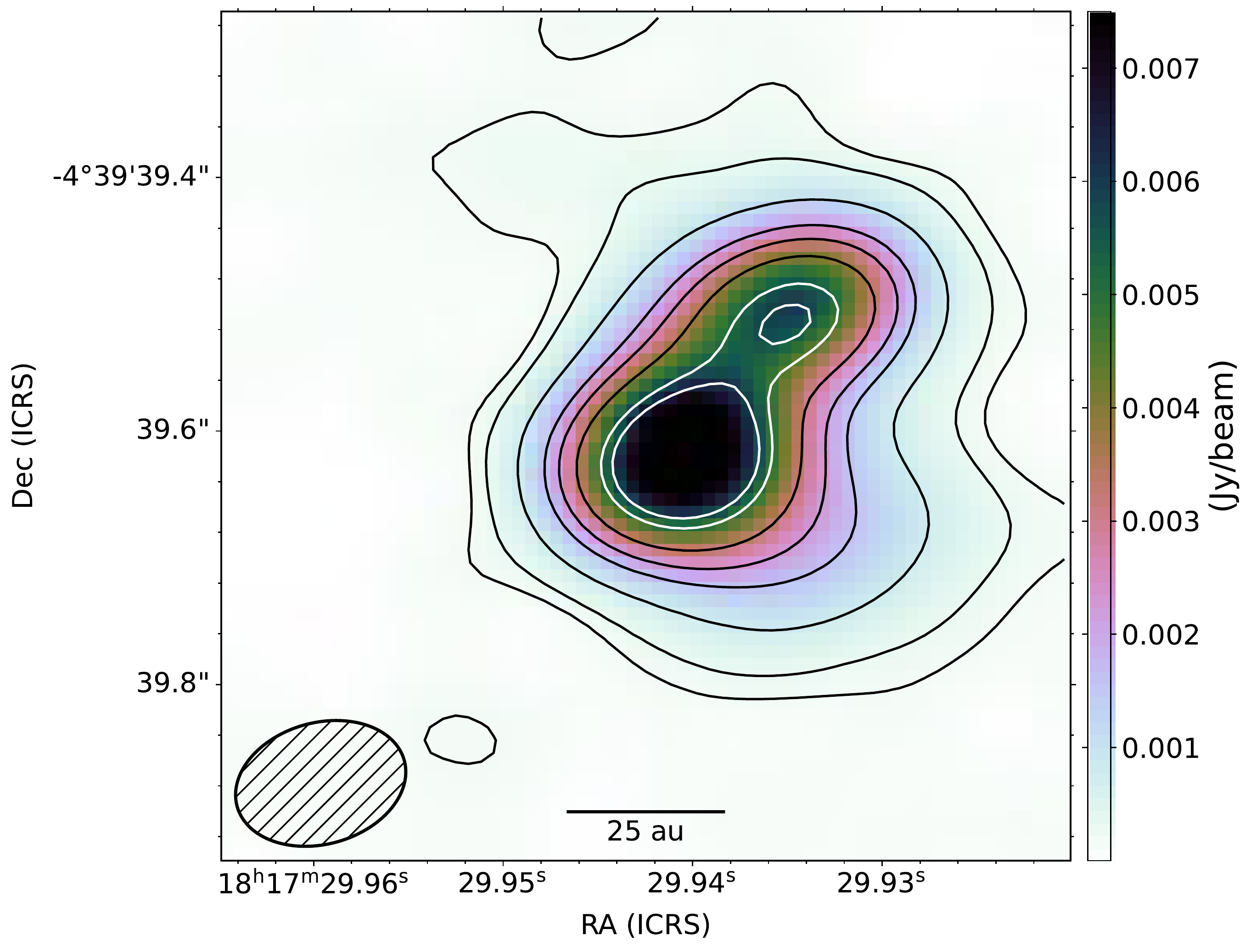}
\caption{ALMA image of L483 using robust = 0.5 weighting.  Black contours at 6, 10, 25, 50, 70, 100$\sigma$, where $\sigma = 33 \mu Jy$/beam. White contours showing 150 and 175$\sigma$.}
\label{fig:briggs}
\end{figure}

\end{document}